\documentclass[prd,nofootinbib,12pt]{revtex4}

\UseRawInputEncoding
 
\usepackage[colorlinks=true,linkcolor=red,citecolor=blue]{hyperref}
\usepackage{amsmath}
\usepackage{amsfonts}
\usepackage{graphicx}
\usepackage{subfigure}
\usepackage{dcolumn}
\usepackage{bm}
\usepackage{booktabs}
\usepackage{multirow}
\usepackage{graphicx,graphics,dcolumn,booktabs,bm}
\usepackage{longtable,lscape}
\usepackage{txfonts}
\usepackage{overpic}
\usepackage{amssymb}
\usepackage{indentfirst}
\usepackage{epsfig}
\usepackage{feynmf}   %{feynmp}
\usepackage{epstopdf}   %{feynmp}
\usepackage{slashed}  %for Feynman symbols
\usepackage{color}
\usepackage[section]{placeins}

\def\haS{\hat{\alpha}_s}
\def\cbar{\overline{c}}

\def\taubar{\overline{\tau}}

\def\Heff{\mathcal{H}_{\rm eff}}

\def\O{\mathcal{O}}
\def\nn{\nonumber}
\begin{document}
\title{\boldmath Revisit to the $b\to c\tau\nu$ transition: in and beyond the SM}
\author{Kingman Cheung$^{1,2,3}$}
\email{cheung@phys.nthu.edu.tw}
\author{Zhuo-Ran Huang$^{4,5}$}
\email{zhuang@lal.in2p3.fr}
\author{Hua-Dong Li$^{5,6}$}
\email{lihd@ihep.ac.cn}
\author{Cai-Dian L\"u$^{5,6}$}
\email{lucd@ihep.ac.cn}
\author{Ying-nan Mao$^1$}
\email{ynmao@cts.nthu.edu.tw}
\author{Ru-Ying Tang$^{5,6}$}
\email{tangruying@ihep.ac.cn}
\affiliation{
$^1$Physics Division, National Center for Theoretical Sciences, Hsinchu, Taiwan 300\\
$^2$Department of Physics, National Tsing Hua University, Hsinchu, Taiwan 300\\
$^3$Division of Quantum Phases and Devices, School of Physics, Konkuk University, Seoul 143-701, Republic of Korea\\
$^4$Universit\'e Paris-Saclay, CNRS/IN2P3, IJCLab, 91405 Orsay, France\\
$^5$Institute of High Energy Physics, Chinese Academy of Sciences, Beijing 100049, China\\
$^6$School of Physics, University of Chinese Academy of Sciences, Beijing 100049, China
}

\begin{abstract}
We perform an analysis of the $b\to c\tau\nu$ data, including $R(D^{(*)})$, $R(J/\psi)$, $P_\tau(D^{*})$ and $F_L^{D^*}$, within and beyond the Standard Model (SM). We fit the $B\to D^{(*)}$ hadronic form factors in the HQET parametrization to the lattice and the light-cone sum rule (LCSR) results, applying the general strong unitarity bounds corresponding to $J^P=1^-$, $1^+$, $0^-$ and $0^+$. Using the obtained HQET relations between helicity amplitudes, we give the strong unitarity bounds on individual helicity amplitudes, which can be used in the BGL fits. Using the fitted form factors and taking into account the most recent Belle measurement of $R(D^{(*)})$ we investigate the model-independent and the leptoquark model explanations of the $b\to c\tau\nu$ anomalies. Specifically, we consider the one-operator, the two-operator new physics (NP) scenarios and the NP models with a single $R_2$, $S_1$ or $U_1$ leptoquark which is supposed to be able to address the $b\to c\tau\nu$ anomalies, and our results show that the $R_2$ leptoquark model is in tension with the limit $\mathcal B(B_c\to \tau\nu)<10\%$. Furthermore, we give predictions for the various observables in the SM and the NP scenarios/leptoquark models based on the present form factor study and the analysis of NP.
\end{abstract}

\maketitle

%================================================================================
%================================================================================
\section{Introduction}
\label{sec:intro}
In recent years several discrepancies between the measurements and the Standard Model (SM) predictions have been observed in the bottom sector, particularly the $R(D^{(*)})$ anomalies in the charged current transition $b\to c\tau\nu$ and the $R_{K^{(*)}}$ anomalies in the neutral current transition $b\to s\ell^+\ell^-$ ($\ell=e,\mu$), which both hint the violation of lepton flavour universality (LFU)\footnote{For reviews, see \cite{Li:2018lxi,Bifani:2018zmi}.}. In view of no existing clear signal of NP from direct searches at the Large Hadron Collider (LHC) and other high energy facilities, these implications of NP at low energy have become quite valuable. Furthermore, these LFU ratios are constructed to be theoretically clean, which means the relevant hadronic uncertainties tend to be cancelled, therefore it is not surprising that they have attracted widespread attention in the high energy physics community. However, the ``full'' cancellation of uncertainties due to hadronic matrix elements is only true for $R_{K^{(*)}}$ but not for the case of $R(D^{(*)})$ which is defined as
\begin{eqnarray}
    R(D^{(*)})=\frac{\mathcal B(B\to D^{(*)}\tau\nu)}{\mathcal B(B\to D^{(*)}\ell\nu)},\quad \text{ with } \ell=\mu, e.\label{eq:RDst}
\end{eqnarray}
In Eq.~\eqref{eq:RDst}, the mass differences between the tauon and the muon/electron lead to the difference in phase spaces between the semi-tauonic and the semi-leptonic decay modes, which result in the partial cancellation of the uncertainties due to the $B\to D^{(*)}$ form factors in the ratios and slight discrepancies in the SM predictions for $R(D)$~\cite{Bigi:2016mdz,Bernlochner:2017jka,Jaiswal:2017rve,Huang:2018nnq,Wang:2017jow,Aebischer:2018iyb,Murgui:2019czp,Zhong:2018exo,Tanaka:2012nw,Hu:2019bdf,Yao:2019vty,Zhou:2019stx,Biancofiore:2013ki,Datta:2012qk,Ivanov:2016qtw,Iguro:2020cpg} and $R(D^*)$~\cite{Bernlochner:2017jka,Jaiswal:2017rve,Bigi:2017jbd,Fajfer:2012vx,Huang:2018nnq,Tanaka:2012nw,Aebischer:2018iyb,Bordone:2019vic,Hu:2019bdf,Zhou:2019stx,Biancofiore:2013ki,Jaiswal:2020wer,Datta:2012qk,Ivanov:2016qtw,Iguro:2020cpg}. Given that the most recent Belle measurement of $R(D^{(*)})$~\cite{Abdesselam:2019dgh,Belle:2019rba}\footnote{For theoretical analysis including the Belle 2019 data, see~\cite{Bardhan:2019ljo,Blanke:2019qrx,Blanke:2018yud,Alok:2019uqc,Shi:2019gxi,Becirevic:2019tpx,Jaiswal:2020wer,Mu:2019bin}.} is consistent with the SM prediction averaged by HFLAV within $1.2\sigma$~\cite{Amhis:2016xyh}, the theoretical determination of $R(D^{(*)})$ in and beyond the SM is still interesting and important.

The existing theoretical estimations of the $B\to D^{(*)}$ form factors have been mostly performed by using the Caprini-Lellouch-Neubert (CLN) parametrization~\cite{Caprini:1997mu}, the Boyd-Grinstein-Lebed (BGL) parametrization~\cite{Boyd:1994tt,Boyd:1997kz} and their variants or updated versions \cite{Bernlochner:2017jka,Jung:2018lfu,Bordone:2019vic,Bourrely:2008za,Gambino:2019sif}. All these parameterizations are based on the analytic properties of the QCD correlation functions, the dispersion relation and the crossing symmetry, with the difference being that in the CLN(-like) parametrizations the heavy quark symmetry is employed to relate the form factors while in the BGL(-like) parametrizations the $B_c$ pole(s) are separated from the analytic parts that are expanded in z and truncated at a certain level. In addition to the original CLN parametrization, the HQET parametrization including the $\mathcal{O}(\alpha_s,\Lambda_{\mathrm{QCD}}/m_{b,c})$ corrections for all $b\to c$ currents has been adopted in \cite{Bernlochner:2017jka} as well as some recent works~\cite{Jung:2018lfu,Aaij:2019ths} to perform global fits.

On the other hand, the deviation between the current world average~\cite{Amhis:2016xyh} of the measurements~\cite{Lees:2012xj,Lees:2013uzd,Huschle:2015rga,Sato:2016svk,Hirose:2016wfn,Aaij:2015yra,Aaij:2017uff,Aaij:2017deq,Abdesselam:2019dgh} of $R(D^{(*)})$ and the SM prediction is still at the level of $3-4\sigma$ depending on the determination of hadronic form factors, therefore it is well worth reexamining the NP explanations of the $b\to c\tau\nu$ anomalies along with the form factor study. To address the $R(D^{(*)})$ anomalies, NP contributions comparable to the tree-level SM prediction are required, which can be satisfied by three categories of NP models: the leptoquark (LQ) models~\cite{Sakaki:2013bfa,Dorsner:2016wpm,Li:2016vvp,Angelescu:2018tyl,Iguro:2018vqb,Fajfer:2015ycq,Becirevic:2016yqi,Crivellin:2017zlb,Yan:2019hpm,Bansal:2018nwp,Altmannshofer:2017poe,Blanke:2018sro,Calibbi:2017qbu,Biswas:2018jun,Bhattacharya:2014wla,Bhattacharya:2016mcc,Hu:2018lmk,Hu:2018veh}, the models with a charged vector boson ($W'$ models)~\cite{Gomez:2019xfw,Faroughy:2016osc,He:2012zp,Greljo:2015mma,Babu:2018vrl,Asadi:2018wea,Cvetic:2017gkt,Boucenna:2016qad,Boucenna:2016wpr} and the charged Higgs models~\cite{Crivellin:2012ye,Crivellin:2015hha,Celis:2012dk,Tanaka:2012nw,Ko:2012sv,Kim:2015zla,Fuyuto:2017sys,Iguro:2018oou,Celis:2016azn,Li:2019xmi,Li:2018rax}.

A LQ is a scalar or a vector boson which carries both lepton number and baryon number and interacts with a lepton and a quark directly. Such particles can arise naturally in some unification models, such as the minimal grand unification [$\textrm{SU}(5)$] model \cite{Georgi:1974sy} and the Pati-Salam [$\textrm{SU}(4)\times\textrm{SU}(2)^2$] model \cite{Pati:1974yy}, etc. Some types of the LQs can form the $bc\tau\nu$ interaction and thus can explain the charged current B anomalies. Like the $W$ boson in the SM, a $W'$ boson which extends the SM gauge sector can also mediate the $b\to c\tau\nu$ transition at the tree level. Such an extra vector boson can appear in models with an enlarged gauge group such as the left-right models \cite{Mohapatra:1974hk}. Besides, charged Higgs bosons have also been considered as a candidate to resolve the $b\to c\tau\nu$ puzzle. A charged Higgs can arise in models with an extended scalar sector, such as the two-Higgs-doublet models \cite{Branco:2011iw}, or the Georgi-Machacek model \cite{Georgi:1985nv} etc.

Among these models, the charged Higgs models are ruled out by the bound from $B_c$ lifetime~\cite{Alonso:2016oyd}, and more stringently from $\mathcal B(B_c\to\tau\nu)$~\cite{Akeroyd:2017mhr}, while the simplified $W'$ models are disfavoured by high-$p_T$ experiments~\cite{Faroughy:2016osc}. Therefore, the LQ models appear to be the most interesting, especially the vector LQ $U_1$ with the SM quantum numbers ($\textbf{3}$, $\textbf{1}$, $\frac{\textbf{2}}{\textbf{3}}$), which is capable of simultaneously addressing $R(D^{(*)})$ and $R_{K^{(*)}}$ \cite{Kumar:2018kmr}. Besides $U_1$, the scalar LQs $R_2$ and $S_1$, respectively with the SM charges ($\textbf{3}$, $\textbf{2}$, $\frac{\textbf{7}}{\textbf{6}}$) and ($\bar{\textbf{3}}$, $\textbf{1}$, $\frac{\textbf{1}}{\textbf{3}}$) are the other two single LQ scenarios which can explain the $b\to c\tau\nu$ anomalies.

In this work, we study the $b\to c\tau\nu$ anomalies by paying efforts to both the determination of the $B\to D^{(*)}$ form factors and the probe of new physics. We perform our fit to recent results calculated in lattice QCD and QCD light-cone sum rules (LCSR) which are valid for complementary $q^2$, following the parametrization adopted in \cite{Jung:2018lfu} and \cite{Huang:2018nnq}, which is based on the HQET parametrization in \cite{Bernlochner:2017jka} but include $\mathcal{O}(\varepsilon_c^2)$ corrections to the form factors of which the $\mathcal{O}(\varepsilon_c)$ corrections are vanishing to avoid a too restrictive parametrization. In the fit, we impose the strong unitarity bounds that follow from the analytic properties of QCD correlation functions and the crossing symmetry. Using the fitted $B\to D^{(*)}$ form factors and taking into account the most recent experimental data, we update the model-independent analysis of the NP effects in the $b\to c\tau\nu$ transition within the framework of the weak effective theory. In our analysis, we consider all the $b\to c\tau\nu$ data, including not only $R(D^{(*)})$, but also the ratio $R(J/\psi)$~\cite{Aaij:2017tyk} which deviate from the SM prediction at the level of $2\sigma$~\cite{Huang:2018nnq}, and the longitudinal polarization fractions of $\tau$ ($P_\tau(D^*)$)~\cite{Hirose:2016wfn} and $D^{*}$ ($F_L^{D^*}$)~\cite{Abdesselam:2019wbt,Adamczyk:2019wyt}. Furthermore, we study the models with a single LQ which can explain the $b\to c\tau\nu$ anomalies, including models with a scalar LQ $R_2$, $S_1$ or a vector LQ $U_1$.

The work is organised as follows: we give a brief introduction to the effective Hamilton relevant to the $b\to cl\nu$ transition in Section~\ref{sec:eff} and then present the fit of the $B\to D^{(*)}$ form factors and the derivation of the unitarity constraints in Section~\ref{sec:ffs}. In Section~\ref{sec:np} we present the study of the model-independent scenarios of NP and the LQ models, and moreover, we give SM and NP predictions of various observables for the $b\to c\tau\nu$ transition. Finally in Section~\ref{sec:con}, we give our summary and conclusions.
\section{Weak effective Hamiltonian}
\label{sec:eff}
The $b\to c\tau\nu$ transition can be described by the weak effective theory. Without the right-handed neutrinos, the effective Hamiltonian can be written as
\begin{equation}
   \Heff = {4G_F \over \sqrt2} V_{cb}\left[ (1 + C_{V_1})\O_{V_1} + C_{V_2}\O_{V_2} + C_{S_1}\O_{S_1} + C_{S_2}\O_{S_2} + C_T\O_T \right] + \text{H.c.} \,,
      \label{eq:lag}
\end{equation}
where the four-fermion operators $\O_X$ ($X=S_1$, $S_2$, $V_1$, $V_2$, and $T$) with different Lorentz and chiral structures form the effective operator basis, and $C_X$ are the corresponding Wilson coefficients. The effective operators are defined as
\begin{eqnarray}
 &\O_{S_1} =  (\cbar_L b_R)(\taubar_R \nu_{L}) \,, \,\,\,
    \O_{S_2} =  (\cbar_R b_L)(\taubar_R \nu_{L}) \,, \nonumber \\
  &  \O_{V_1} = (\cbar_L \gamma^\mu b_L)(\taubar_L \gamma_\mu \nu_{L}) \,,  \,\,\,
     \O_{V_2} = (\cbar_R \gamma^\mu b_R)(\taubar_L \gamma_\mu \nu_{L}) \,, \nonumber \\
  & \O_T =  (\cbar_R \sigma^{\mu\nu} b_L)(\taubar_R \sigma_{\mu\nu} \nu_{L}) \,,
   \label{eq:operators}
\end{eqnarray}
where $\O_{V_1}$ is the only operator present in the SM and the Wilson coefficients of $\O_X$ get modified by the NP at the higher energy scale. Starting from the effective Hamiltonian given in Eq.~\eqref{eq:lag}, the observables involved in this work, i.e., $R(D^{(*)})$, $R(J/\psi)$, $P_\tau(D^*)$ and $F_L^{D^*}$ can be computed and expressed in terms of $C_X$, which of course require the information of the hadronic matrix elements. For the $B_c\to J/\psi$ channel, we use the form factors calculated in the covariant light-front quark model~\cite{Wang:2008xt}, which have been shown in \cite{Huang:2018nnq} the consistency with the preliminary lattice QCD results~\cite{Lytle:2016ixw,Colquhoun:2016osw}. Regarding the $B\to D^{(*)}$ form factors, we perform a fit in the next section.
\section{$B\to D^{(*)}$ form factors}
\label{sec:ffs}
Most of the recent determinations of the $B\to D^{(*)}$ form factors are based on the HQET parametrizations (including the CLN parametrization~\cite{Caprini:1997mu} and the more recent versions~\cite{Bernlochner:2017jka,Jung:2018lfu}), and the BGL~\cite{Boyd:1994tt,Boyd:1997kz} (and the BCL~\cite{Bourrely:2008za}) parametrizations. The main difference between these two classes of parametrizations is that for the latters the heavy quark symmetry is not employed. Such kind of parametrizations leave the form factors maximally model-independent but on the other hand their validity also rely more on the inputs. In this work, we follow the parametrization proposed in \cite{Jung:2018lfu}, which include in HQET parametrization the next-to-leading order contributions in $\alpha_s$ and $\Lambda_{\mathrm{QCD}}/m_{b,c}$, and the $\mathcal{O}(\varepsilon_c^2)$ corrections to the form factors which do not have $\Lambda_{\mathrm{QCD}}/m_{b,c}$ corrections in order to avoid a too restrictive parametrization. To obtain reliable predictions for the $B\to D^{(*)}$ form factors, we consider the data points calculated in the QCD-based approaches including lattice QCD and QCD light-cone sum rules. These two methods provide us information on the $B\to D^{(*)}$ form factors in complementary kinematical regions and thus can be used to determine the form factors in the full $q^2$ region. Furthermore, following the analytic properties of the QCD correlation functions and the crossing symmetry, the $B\to D^{(*)}$ form factors should fulfill the unitarity constraints, which have been considered in some existing works~\cite{Bigi:2016mdz,Bigi:2017jbd,Bigi:2017njr,Jung:2018lfu}. In our fit, we employ the strong unitarity bounds on helicity amplitudes corresponding to quantum numbers $J^P=1^-/1^+/0^-/0^+$. Furthermore, using the heavy quark symmetry relations (including the effects due to the higher-order terms in $\Lambda_{\mathrm{QCD}}/m_{b,c}$) between the $B^{(*)}\to D^{(*)}$ helicity amplitudes, we derive the strong bounds for the individual helicity amplitudes to check the consistency of the fit.
\subsection{Fit with the HQET parametrization}
\label{subsec:fit}
To illustrate the typical structure of the parametrization we adopt for the $B\to D^{(*)}$ form factors of different currents, we write explicitly the $B\to D$ hadronic matrix element of the vector current $\bar c\gamma^\mu b$ and the relevant form factors $h_+(w)$ and $h_-(w)$:\footnote{The full expressions for all $B\to D^{(*)}$ form factors can be found in \cite{Huang:2018nnq} and with greater details in \cite{Bernlochner:2017jka}. We use the input parameters such as the b(c) quark masses, the $B(D^{(*)})$ meson masses, the QCD coupling etc. following \cite{Huang:2018nnq}.}
\begin{eqnarray}
\langle D(k)| \bar c\gamma^\mu b |\overline B(p)\rangle &=& \sqrt{m_Bm_D} \left [ h_+(w) (v+v')^\mu + h_-(w) (v-v')^\mu \right ], \label{eq:BtoDffhqet}\\
 h_+&=&\xi\bigg\{1 + \haS\Big[\hat C_{V_1} + \frac{w+1}2\, (\hat C_{V_2}+\hat C_{V_3})\Big]
  + (\varepsilon_c + \varepsilon_b)\, \hat{L}_1\bigg\},\\
 h_-&=& \xi\big[\haS\, \frac{w+1}2\, (\hat C_{V_2}-\hat C_{V_3}) + (\varepsilon_c - \varepsilon_b)\, \hat{L}_4\big],
\end{eqnarray}
where HQET allows us to express the $B\to D^{(*)}$ form factors in terms of a single leading Isgur-Wise function $\xi(w)$ in the heavy quark limit, which is normalized to 1 at the zero hadronic recoil. As in \cite{Jung:2018lfu}, the extrapolation of $\xi(w)$ is kept up to the $\mathcal O(z^2)$ corrections in the $z$ expansion, i.e., $\xi(z) = 1 - 8  \rho^2  z + (64  c - 16  \rho^2)z^2$ is used with $\rho^2$ and $c$ to be determined in the fit, and $z(w) = (\sqrt{w+1}- \sqrt{2})/(\sqrt{w+1} + \sqrt{2})$ with $w=\frac{m_{B}^2+m^2_{D^{(*)}}-q^2}{2m_{B}m_{D^{(*)}}}$ is the conformal mapping of the $q^2$ plane to a unit disk. The $\mathcal{O}(\varepsilon_{b,c})$ (where $\varepsilon_{b,c}={\bar\Lambda\over 2m_{b,c}}$ with $\bar\Lambda\sim\mathcal O(\Lambda_{QCD})$~\cite{Huang:2018nnq}) contributions in the heavy quark expansion are represented by the functions $\hat{L}_{1\ldots6}$ that can be expressed in terms of the sub-leading Isgur-Wise functions $\hat \chi_2(w)$, $\hat \chi_3(w)$ and $\eta(w)$, which can be expanded at $w=1$ and give us 5 fitting parameters\footnote{The Luke's theorem implies $\hat{\chi}_3(1) = 0$.} up to $\mathcal{O}(\varepsilon_{c,b}(w-1))$. Furthermore, three parameters $\delta_{h_{A_1}}$, $\delta_{h_+}$ and $\delta_{h_{A_1}}$ are the coefficients of the $\mathcal{O}(\varepsilon_c^2)$ corrections, while the coefficients of the $\mathcal O(\alpha_s)$ corrections expressed by $\hat C_X(w,\frac{m_c}{m_b})$ ($X=S$, $P$, $V_i$, $A_i$ and $T_i$) given in \cite{Bernlochner:2017jka} do not involve any free parameters. Therefore altogether there are 10 parameters to be fitted.

In the fit, we use data points calculated in lattice QCD and QCD light-cone sum rules, which are valid in complementary kinematical regions. For the $B\to D$ form factors, the HPQCD and the Fermilab/MILC Collaborations have performed unquenched lattice QCD calculations. HPQCD has provided z-expansion in the BCL parametrization, for which the expansion coefficients and the corresponding covariance matrices are given \cite{Na:2015kha}. Fermilab/MILC has given results of $f_+(w)$ and $f_0(w)$ at $w=1$, $1.08$ and $1.16$ along with the correlations in \cite{Lattice:2015rga}. Moreover, the FLAG Collaboration has performed a combined fit of the $B\to D^{*}$ form factors using the BCL parametrization, which we do not take in this work but instead use the original results obtained by HPQCD and Fermilab/MILC.\footnote{However we take the same data points as FLAG at $w=1$ and 1.1 from the z-expansion given by HPQCD.} For the $B\to D^*$ transition, we take the result $\mathcal{F}^{B\to D^*}\!(1) = h_{A_1}\!(1) = 0.904(12)$ obtained in \cite{Harrison:2017fmw}, which is the average value of the results at the zero hadronic recoil calculated by Fermilab/MILC \cite{Bailey:2014tva} and HPQCD \cite{Harrison:2017fmw}.\footnote{There are also preliminary results of the $B\to D^{*}$ form factors for non-zero hadronic recoil respectively by the JLQCD \cite{Kaneko:2018mcr} and Fermilab/MILC \cite{Vaquero:2019ary,Aviles-Casco:2019vin} collaborations, however in this work we only use the finalized results.} In total, we use 11 data points from lattice QCD.

QCD light-cone sum rule is another QCD-based approach to estimate the $B\to D^{(*)}$ form factors. The starting point of LCSR is the meson-to-vacuum two-point correlation function. The dispersion relation allows to match the QCD representation to the phenomenological representation of the correlation function via the quark-hadron duality. At the QCD side, the correlation function is expanded near the light-cone $x^2=0$ in terms of the meson distribution amplitudes (DAs) that encode the non-perturbative QCD effects. Since the light-cone dominance of the correlation function is satisfied for small momentum transfer $q^2$ in $B\to D^{(*)}$ transitions \cite{Faller:2008tr}, LCSR is a valuable method to compute the form factors in the large hadronic recoil region. In our analysis, we use the LCSR results calculated using the B meson DAs in \cite{Faller:2008tr} up to leading order (LO) in the $\alpha_s$ and the light-cone expansion, \cite{Wang:2017jow} up to next-to-leading order in the $\alpha_s$ expansion and \cite{Gubernari:2018wyi} up to twist-4 two-particle and three-particle contributions in the light-cone expansion. In our fit, we take the $B\to D^{(*)}$ form factors in \cite{Faller:2008tr,Wang:2017jow,Gubernari:2018wyi} for the maximal hadronic recoil. In addition we extrapolate the results in \cite{Faller:2008tr,Wang:2017jow} to $w=1.4$ and $1.3$ using the slope parameters given there in order to constrain the shapes of the form factors in the large hadronic recoil region.\footnote{In \cite{Gubernari:2018wyi} form factors at several points of negative $q^2$ and $q^2=5$ have been given, however, in this work we don't take any data points at negative $q^2$, and we also neglect the results at $q^2=5$ in \cite{Gubernari:2018wyi} because those results are associated with large ($50\%$ or more) contributions from the twist-4 two-particle DAs. This may also imply sizable contributions from twist-4 multi-particle DAs or even higher twist DAs, particularly for $q^2\geqq5$.} Therefore, altogether we have 22 LCSR data points in the fit.
\subsection{Weak unitarity bounds}
As mentioned in the previous sections, the $B\to D^{(*)}$ form factors should fulfill the model-independent unitarity bounds following the analytic properties of the QCD correlators and the crossing symmetry. To introduce the unitarity bounds, we still write the hadronic matrix elements for the $B\to D$ transition as in \eqref{eq:BtoDffhqet}, but in the general form
\begin{align}
\langle D(k)| \bar c\gamma^\mu b |\bar{B}(p)\rangle&=f_+(q^{2})(p+k)^{\mu}
+f_{-}(q^{2})(p-k)^{\mu},
\end{align}
where $q=p-k$ and $f_-(q^2)$ can be rewritten as $f_-(q^2)=\frac{m_B^2-m_D^2}{q^2}(f_0(q^2)-f_+(q^2))$. The form factors $f_{+,0}$ can be parameterized as
\begin{align}
\label{eq:fpf0}
f_{+,0}(z)=\frac{1}{P_{+,0}(z) \phi_{+,0}(z)}\sum_{n=0}^{\infty}a_{(f_+,f_0)n} z^{n}(w),
\end{align}
where $P_{+,0}(z)$ and $\phi_{+,0}(z)$ are respectively termed as the Blaschke factors and the outer functions. The Blaschke factors absorb the $B_c$ resonances below the $BD$ pair production threshold $t_+=(m_B+m_D)^2$, ensuring the proper analytic behaviour of the form factors. The explicit expressions of the Blaschke factors are
\begin{align}\label{blaschke}
P_{+,0}(z)=\prod_{P_{+,0}=1}^{j}\frac{z-z_{P_{+,0}}}{1-zz_{P_{+,0}}},
\end{align}
with $j=3$ for $f_+$ and $j=2$ for $f_0$, and $z_{P}$ defined as:
\begin{equation*}
z_{P}=\frac{t_--m_p^2}{(\sqrt{t_+-m_p^2}+\sqrt{t_+-t_-})^2},
\end{equation*}
where $t_-=(m_B-m_D)^2$, and $m_p$ refer to the masses of the $B_c$ narrow resonances, of which we use the most recent results obtained in experiments, lattice QCD and the nonrelativistic quark model summarized in Table~\ref{tab:Bcmass}. The outer functions corresponding to different form factors are given in~\cite{Boyd:1997kz}.
\begin{table*}[!htbp]\small%
	\centering
	\begin{tabular}{ccccccc}
		\hline
		\hline
		$B_{c}(1\ S^{\!\!\!\!\!\!\!1}_{0})$&$B_{c}(2\ S^{\!\!\!\!\!\!\!1}_{0})$&$B_{c}(2\ S^{\!\!\!\!\!\!\!3}_{1})$&$B_{c}(1\ S^{\!\!\!\!\!\!\!3}_{1})$&$B_{c}(1\ P^{\!\!\!\!\!\!\!\!3}_{0})$&$B_{c}(1P_{1})$&\\
		\hline
		6.271  & 6.871 &6.840  & 6.331 & 6.712 &6.736 &\\
		\hline
			\multicolumn{3}{c}{Experiments \cite{CMSBc:2019,LHCbBc:2019}}&\multicolumn{3}{c}{Lattice QCD \cite{Nilmani:2018}}&\\
		\hline
		\hline
			$B_{c}(1\ D^{\!\!\!\!\!\!\!\!3}_{1})$&	$B_{c}(3\ S^{\!\!\!\!\!\!\!3}_{1})$&$B_{c}(2\ P^{\!\!\!\!\!\!\!\!3}_{0})$&	$B_{c}(3\ S^{\!\!\!\!\!\!\!1}_{0})$&$B_{c}(1P'_{1})$&$B_{c}(2P_{1})$&$B_{c}(2P'_{1})$\\
		\hline
		7.020&7.252 & 7.107  &7.239&6.776&7.134 & 7.150 \\
		\hline
		\multicolumn{7}{c}{Nonrelativistic quark model \cite{Li:2019}}\\
		\hline
		\hline
	\end{tabular}	
	\caption{Masses of $ B_{c} $ in units of GeV.}
	\label{tab:Bcmass}
\end{table*}
The core element for deriving the unitarity constraints is the two-point QCD correlation function of local composite operators. For the vector current $J^{\mu}=\bar c\gamma^\mu b$, the two-point correlation function can be written as
\begin{align}
\label{correlation}
i\int d^{4}xe^{iqx}\langle0\mid TJ^{\mu}(x)J^{\nu}(0)\mid0\rangle=\big[(q^{\mu}q^{\nu}-q^{2}g^{\mu\nu})\Pi^{T}(q^{2})+q^{\mu}q^{\nu}\Pi^{L}(q^{2})\big]/q^{2},
\end{align}
where the invariants $\Pi^{T}(q^2)$ and $\Pi^{L}(q^2)$ correspond to $J^P=1^-$ and $0^+$ respectively, which satisfy the standard dispersion relation
\begin{align}\label{eq:disper}
\Pi^{T,L}(q^{2})=\frac{1}{\pi}\int_{0}^{\infty}dt\frac{Im\Pi^{T,L}(t)}{t-q^{2}-i\epsilon},
\end{align}
Practically, one or more subtractions are necessary to make the dispersion integral finite \cite{Boyd:1997kz}. Inserting a set of two-body intermediate states into the correlation function~(\ref{correlation}) and using the relation Eq.~\eqref{eq:disper} and the crossing symmetry, one obtains the inequality with form factor expressed by $F(t)$\cite{Boyd:1997kz}
\begin{align}\label{eq:ineq}
\frac{1}{\pi\chi^{T,L}(q^{2})}\int_{t_{+}}^{\infty}ds\mid{\widetilde{\phi}}_{+,0}^{T,L}(t,t_-)P_{+,0}(t)F(t)\mid^2 \leqslant1,
\end{align}
where $\chi^{L}=\frac{\partial \Pi^{L}}{\partial q^2}$ and $\chi^{T}=\frac{1}{2}\frac{\partial^2 \Pi^{L}}{\partial (q^2)^2}$ are the subtracted invariant functions, $P_{+,0}(t)$ are the Blaschke factors, and the outer functions $\phi$ are embedded in $\widetilde{\phi}(t,t_-)$. $\chi^{T,L}(q^2)$ can be calculated perturbatively using the operator product expansion (OPE), and in this work we use the results of $\chi(0)$ in \cite{Bigi:2016mdz}, which include the $\mathcal O(\alpha_s^2)$ corrections calculated in \cite{Grigo:2012ji} and neglect the small contribution from the condensate terms. In Eq.~\eqref{eq:ineq}, the hadronic representation of the dispersion integral is less than the subtracted invariant correlator because the inserted hadronic states are only a partial set of the states coupled to the employed interpolating current. Substituting \eqref{eq:fpf0} into \eqref{eq:ineq}, one obtains the (weak) unitarity bounds on the coefficients $a_{(f_+,f_0)n}$. Such bounds read\footnote{Following the same procedure, such constraints can be generalized to any helicity amplitude $F_i$ which is a linear combination of the general form factors and correspond to a certain $J^P$ quantum number.}
\begin{align}
\sum_{n=0}^\infty a_{in}^2<1.
\label{eq:weakbound}
\end{align}
\subsection{Strong unitarity bounds}
The weak bounds Eq.~(\ref{eq:weakbound}) are obtained by considering only a single ($BD$) helicity amplitude in \eqref{eq:ineq}. If we
consider all $B^{(*)}D^{(*)}$ intermediate states corresponding to certain quantum numbers, we can get the strong unitarity bounds. We take the $1^-$ quantum number as an example to illustrate the derivation of the strong unitarity bounds. Considering all the $B^{(*)}D^{(*)}$ intermediate states, there are 7 helicity amplitudes corresponding to $J^P=1^-$ listed in Table~\ref{tab:HA}. Here we adopt the notations in \cite{Boyd:1997kz} for the helicity amplitudes and their definitions can also be found therein.
\renewcommand{\arraystretch}{1.3} % default is 1.0
\begin{table*}[!htbp]\small%
	\begin{center}
		\begin{tabular}{|c|c|c|c|}
			\hline
			$B\to D$ &$B\to D^*$ & $B^*\to D$ & $B^*\to D^*$ \\
			\hline
			$f_+$ & $ g$ & $\hat{g}$ & $V_{+0}$, $V_{++}$, $V_{0+}$, $V_{00}$\\
			\hline
		\end{tabular}%\end{tabular*}
	\end{center}
	\caption{Vector helicity amplitudes in notations of~\cite{Boyd:1997kz}.\label{tab:HA}}
\end{table*}

Analogous to Eq.~\eqref{eq:fpf0}, these helicity amplitudes can also be parameterized as~\cite{Boyd:1997kz, Bigi:2016mdz}
\begin{align}
\label{eq:strong1}
F_{i}=\widetilde{f_{i}}\sum_{n}b_{in}z^{n},
\end{align}
where $F_i$ are respectively $F_1=f_+$, $F_2=g$ ..., and $F_7=V_{00}$, and $\widetilde{f_{i}}$ can be expressed in terms of the corresponding Blaschke factors and outer functions as
\begin{align}
\label{eq:ffrelation}
\widetilde{f_{1}}=\frac{1}{P_{f_+}(z)\phi_{f_+}(z)}, ~~
\widetilde{f_{2}}=\frac{1}{P_{g}(z)\phi_{g}(z)}, ~~
...~~
\widetilde{f_7}=\frac{1}{P_{V_{00}}(z)\phi_{V_{00}}(z)}.\nn \\
\end{align}
Substituting Eq.~\eqref{eq:ffrelation} and \eqref{eq:strong1} into Eq.\eqref{eq:ineq} one obtains the strong unitarity bound
\begin{align}
\label{eq:strong3}
\sum_{i=1}^7 \sum_{n=0}^{\infty} b_{1^-,in}^2\leq 1.
\end{align}
Similarly, one can derive the strong unitarity bounds for other quantum numbers including $1^+$, $0^-$ and $0^+$, which respectively have 7, 7 and 3 corresponding helicity amplitudes. The definitions of these relevant helicity amplitudes can also be found in \cite{Boyd:1997kz} and are not shown here. The strong bounds read
\begin{align}
\label{eq:strong4}
\nonumber\sum_{i=1}^7 \sum_{n=0}^{\infty} b_{1^+,in}^2\leqslant 1,\\
\sum_{i=1}^3 \sum_{n=0}^{\infty} b_{0^-,in}^2\leqslant 1,\\ \nonumber\sum_{i=1}^3 \sum_{n=0}^{\infty} b_{0^+,in}^2\leqslant 1.
\end{align}
\subsection{Results of the fit and strong bounds on individual helicity amplitudes}
Imposing the four strong unitarity bounds \eqref{eq:strong3} and \eqref{eq:strong4} with maximally $n=2$ in the fit of the HQET parametrization to the data points obtained in lattice QCD and LCSR in complementary kinematical regions, we obtain the parameters for the $B\to D^{(*)}$ form factors in Table~\ref{tab:fit_para} with $\chi^2/d.o.f=22.30/23$.
\begin{table}[!htbp]\small%
	\caption{Fitted values of the parameters in the $B\to D^{(*)}$ form factors.}
	\label{tab:fit_para}
	\begin{center}
		\begin{tabular}{ccccc}
			\hline\hline
			$\chi_2(1)$ & $\chi_2'(1)$ & $\chi_3'(1)$  & $\eta(1)$ & $\eta'(1)$\\
			\hline
			0.133(23)  &-0.149(19) &0.017(8) & 0.365(28) & 0.239(114)\\
			\hline\hline
			$\rho^2$ &           c &        $\delta_{h_{A_1}}$ & $\delta_{h_+}$ & $\delta_{h_{T_1}}$\\
			\hline
			1.120(28)  &   0.932(212)     &    -1.304(202)    &   0.032(133)  &  -4.888(1975)  \\
			\hline
			\hline
		\end{tabular}
	\end{center}
\end{table}

Furthermore, using the heavy quark relations between the helicity amplitudes with the same quantum number, for example, with $J^P=1^-$, the expanding coefficients of the helicity amplitudes can be related through
\begin{align}
\label{eq:ffsrelation}
\sum_n a_{f_+n} z^n \simeq& \sum_n b_{f_+n}z^n, \nn \\
\sum_n a_{f_+n} z^n \simeq& \frac{ P_{f_+}(z)\phi_{f_+}(z)f_+}{P_g(z)\phi_g(z)
	g}\sum_n b_{gn}z^n, \nn \\
... \nn \\
\sum_n a_{f_+n} z^n \simeq& \frac{ P_{f_+}(z)\phi_{f_+}(z)f_+}{P_{V_{00}}(z)\phi_{V_{00}}(z)
	V_{00}}\sum_n b_{V_{00}n}z^n,
\end{align}
which allows us to express the strong unitarity constraints in terms of only $a_{f_+n}$. For the other quantum numbers, analogous relations also hold. The derivation of such constraints on expanding coefficients of a single helicity amplitude requires the HQET relations in terms of $(w-1)$ for the ratios of form factors (e.g. $f_+/g$) on the r.h.s of Eq.~\eqref{eq:ffsrelation}, which depend on the HQET parameters that have been obtained in Table~\ref{tab:fit_para}. Based on these relations we further obtain the following strong unitarity bounds for individual helicity amplitudes, which can be written in a general form for all quantum numbers as below
\begin{align}
\label{eq:strong5}
\sum_{n=0}^{2}\sum_{m=0}^{n}c_{imn}a_{im}a_{in}\leqslant 1,
\end{align}
where i denotes the helicity amplitudes $ f_{+} $, $ f_{0} $, $ f $ and $ \mathcal{F}_{2} $ \cite{Boyd:1997kz} corresponding to $J^P=1^-$, $0^+$, $1^+$ and $0^-$, and the coefficients $ c_{imn} $ can be found in Table~\ref{tab:coeff}.\footnote{The $1^-$ bound is analogous to the one in \cite{Bigi:2016mdz} obtained by using the HQET parameters in \cite{Bernlochner:2017jka}.}
\begin{table*}[!htbp]\small%
	\begin{center}
		\begin{tabular}{cccc||cccc}
			\hline
			\hline
			$ J^{P}=1^{-} $& $ n=0 $       & $ n=1 $       & $ n=2 $&$ J^{P}=0^{+} $& $ n=0 $       & $ n=1 $       & $ n=2 $       \\
			\hline
			$ m=0  $        & 362.667 &$ -187.514 $ & $ -101.406 $&& 33485.5 & $ -23650.9 $ &$  -7911.52  $\\
			$ m=1  $        &         & 161.785  & 80.4558& &         & 8764.12  & 4390.13  \\
			$ m=2  $        &         &         & 15.58& &         &         & 648.243   \\
			\hline
			\hline
			$ J^{P}=1^{+} $&       &      & &$ J^{P}=0^{-} $&        &        &      \\
			\hline
			$ m=0  $        &15203.5 &$ -8321.63 $ & 3397.97&& 295.779 & $ -121.779 $ &50.4193\\
			$ m=1  $        &         & 1361.13  & $ -965.496 $& &         & 15.313  & $ -9.57116 $  \\
			$ m=2  $        &         &         & 220.067& &         &         & 3.99978  \\
			\hline
			\hline
		\end{tabular}
	\end{center}
	\caption{Coefficients $ c_{imn} $ in the strong unitarity bounds \eqref{eq:strong5} for $i=f_{+}$, $ f_{0} $, $ f $ and $ \mathcal{F}_{2} $. \label{tab:coeff}}
\end{table*}
We check that our fit results also fulfill the bounds \eqref{eq:strong5} which shows the consistency of our fit by imposing the strong unitarity bounds for the HQET parametrization. Moreover, the bounds \eqref{eq:strong5} can be useful for the fitting using BGL parametrization where the helicity amplitudes are not related under heavy quark expansion thus only \eqref{eq:strong5} can be imposed instead of \eqref{eq:strong3} and \eqref{eq:strong4} when only a partial set of $b\to c$ channels such as $B\to D^{(*)}$ are considered.
\section{Numerical analysis}
\label{sec:np}
In this section, we investigate the NP solutions to the $b\to c\tau\nu$ anomalies and compare the SM and NP predictions by using the $B\to D^{(*)}$ form factors determined in the previous section. In our analysis, we take into account the results on $R(D^{(*)})$ reported by BaBar, Belle and LHCb, $R(J/\psi)$ by LHCb, and the longitudinal polarization fractions $P_\tau(D^*)$ and $F_L^{D^*}$ by Belle. All the data and references are collected in Table~\ref{tab:exdata}.\footnote{It should be noted that in some of these analyses, e.g. those on B-factories \cite{Lees:2013uzd,Huschle:2015rga}, the signal shapes will affect on the final results and induce some hidden systematic uncertainties which may be taken for NP. However, in the following numerical analysis as well as analogous analyses in the literature, such possible bias are neglected.} Besides, we also take into account the bound $\mathcal B(B_c\to\tau\nu)<10\%$ from LEP1 data~\cite{Akeroyd:2017mhr} which is more stringent than that from $B_c$ lifetime~\cite{Alonso:2016oyd}. In the next subsections, we give analyses of these data and bounds for both the model-independent NP scenarios and the LQ models.
\begin{table*}[!htbp]\small%
\centering
\caption{Experimental data used in the fits.}
\begin{tabular}{c c c c c }
\hline
\hline
    & $R_D$ & $R_{D^*}$& Correlation & $P_\tau(D^*)$\\
\hline
BaBar\cite{Lees:2012xj, Lees:2013uzd}       &$0.440(58)(42)$    &$0.332(24)(18)$   &$-0.27$  &$-$\\
Belle\cite{Huschle:2015rga}       &$0.375(64)(26)$    &$0.293(38)(15)$   &$-0.49$ &$-$ \\
Belle \cite{Sato:2016svk}      &$-$    &$0.302(30)(11)$  &$-$&$-$ \\
Belle\cite{Hirose:2016wfn}       &$-$    &$0.270(35)(_{-0.025}^{+0.028})$   &$0.33$ &$-0.38(51)(_{-0.16}^{+0.21})$ \\
LHCb  \cite{Aaij:2015yra}       &$-$    &$0.336(27)(30)$   &$-$  &$-$\\
LHCb  \cite{Aaij:2017uff, Aaij:2017deq}       &$-$    &$0.291(19)(26)(13)$   &$-$ &$-$\\
Belle\cite{Abdesselam:2019dgh}   &$0.307(37)(16)$    &$0.283(18)(14)$   &$-0.54$ &$-$\\
\hline
\hline
\end{tabular}
\centering
\begin{tabular}{c c c }
\hline
\hline
     &$R_{J/\psi}$ & $F_L^{D^{*}}$\\
\hline
LHCb\cite{Aaij:2017tyk}       &$0.71(17)(18)$ & $-$\\
Belle\cite{Adamczyk:2019wyt,Abdesselam:2019wbt}& $-$& $0.60(8)(4)$\\
\hline
\hline
\end{tabular}
\label{tab:exdata}
\end{table*}
\subsection{Model-independent scenarios}
Effective field theory (EFT) is a useful tool to deal with NP. It allows to perform global fits to Wilson coefficients of the effective operators without knowing the specific type of new physics at the high energy scale. Such a means of studying NP is model-independent. In our fits, we define the $\chi^2$ as follows
\begin{align}
\chi^2(C_X)=&\sum_{m,n=1}^{\text{data}} (O^{th}(C_X)-O^{exp})_m(V^{exp}+V^{th})_{mn}^{-1} (O^{th}(C_X)-O^{exp})_n \nonumber\\ &+\frac{(R_{J/\psi}^{th}(C_X)-R_{J/\psi}^{exp})^2}{\sigma_{R_{J/\psi}}^2}
+\frac{({F_L^{D^{*}}}^{th}(C_X)-{F_L^{D^{*}}}^{exp})^2}{\sigma_{F_L^{D^{*}}}^2}.
\end{align}
As introduced in Section~\ref{sec:eff}, the effective operators which can induce the $B\rightarrow D^{(*)}\tau\nu$ decay are
\begin{equation}
\mathcal{O}_{S_1},\mathcal{O}_{S_2},\mathcal{O}_{V_1},\mathcal{O}_{V_2},\mathcal{O}_{T}.
\end{equation}
$\mathcal{O}_{V_2}$ can be generated minimally at dimension-8 in the Standard Model Effective Field Theory (SMEFT), while the other four operators can be generated at dimension-6.\footnote{This is because $\mathcal{O}_{V_2}$ is a component of an $\textrm{SU}(2)$ triplet. It can form an $\textrm{SU}(2)$ singlet after at least inserting the Higgs doublet twice. Thus it can be generated minimally at dimension-8 in the SMEFT. While all the other four operators can be extracted directly from an $\textrm{SU}(2)$ singlet, which means all of them can be generated at dimension-6 in the SMEFT.} Thus it means if new physics appear at a high scale $\Lambda$ and the new physics couplings are weak, the Wilson coefficients are naturally expected as $C_{V_1,S_1,S_2,T}\sim\mathcal{O}(v^2/\Lambda^2)$ and $C_{V_2}\sim\mathcal{O}(v^4/\Lambda^4)$.

Assuming new physics induces a single operator with a complex Wilson coefficient or two operators with real Wilson coefficients, we perform global $\chi^2$ fits to the data in Table~\ref{tab:exdata}. The results for all scenarios with and without imposing $\mathcal B(B_c\to\tau\nu)<10\%$ are listed in Table~\ref{tab:wcoef}. Besides, to supplement the global fits, we also give plots which show regions of the Wilson coefficients allowed by different measurements within $2\sigma$ in Figure~\ref{fig:constraint2}and Figures~\ref{fig:constraint2sce1} and \ref{fig:constraint2sce2} in Appendix~\ref{app:A}.

For single-operator scenarios, we observe that the pure scalar scenarios $S_1$ and $S_2$ are already excluded at $2\sigma$ given the $\chi^2$ is greater than 19.7 which is the $95\%$ C.L. excluded limit for $d.o.f=11$, while the other three types are still allowed. The $S_1$ scenario is excluded mainly because it cannot explain $R_D$ and $R_{D^*}$ anomalies simultaneously, as shown in Figure~\ref{fig:constraint2}. Being different from the $S_1$ scenario, the $S_2$ scenario is excluded because of the constraint $\mathcal{B}(B_c\to\tau\nu)<10\%$. The results of the scalar scenarios still disfavour the charged Higgs models as found previously~\cite{Huang:2018nnq,Tran:2018kuv}. For the $T$ scenario, we have $\chi^2_{\textrm{min}}<19.7$ which means it is not excluded at $95\%$ C.L.. But this scenario has a relatively large $\chi^2$ compared with other favoured scenarios because it only accommodates small $F_L^{D^*}$. In Figure~\ref{fig:constraint2}, one can see that the $2\sigma$ allowed region by $F_L^{D^*}$ of the $T$ scenario is small, and the best fit points of this scenario locate outside this region. The vector scenarios $V_1$ and $V_2$ are both favoured by current data at $95\%$ C.L., and it is notable that the global-fit result implies large CP-violation phase for the $V_2$ scenario. In general, the situation for different one-operator NP scenarios is more or less similar to a previous study in \cite{Huang:2018nnq} without the most recent measurements of $F_L^{D^*}$ and $R(D^{(*)})$ by Belle, and with the major notable difference being that $F_L^{D^*}$ poses a relatively strong constraint on the $T$ scenario.
\begin{figure}[!htbp]
\begin{center}
\includegraphics[scale=0.22]{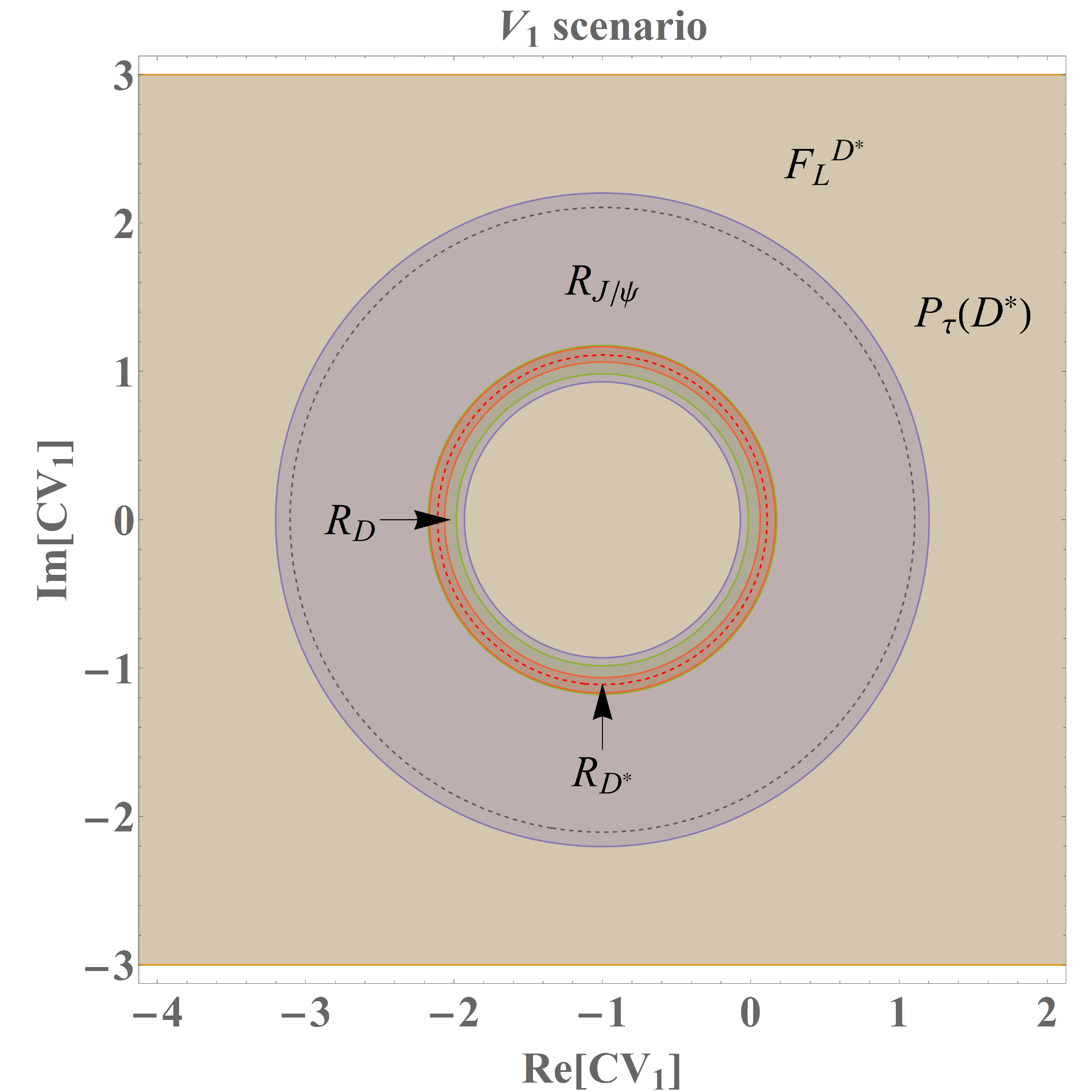}
\includegraphics[scale=0.22]{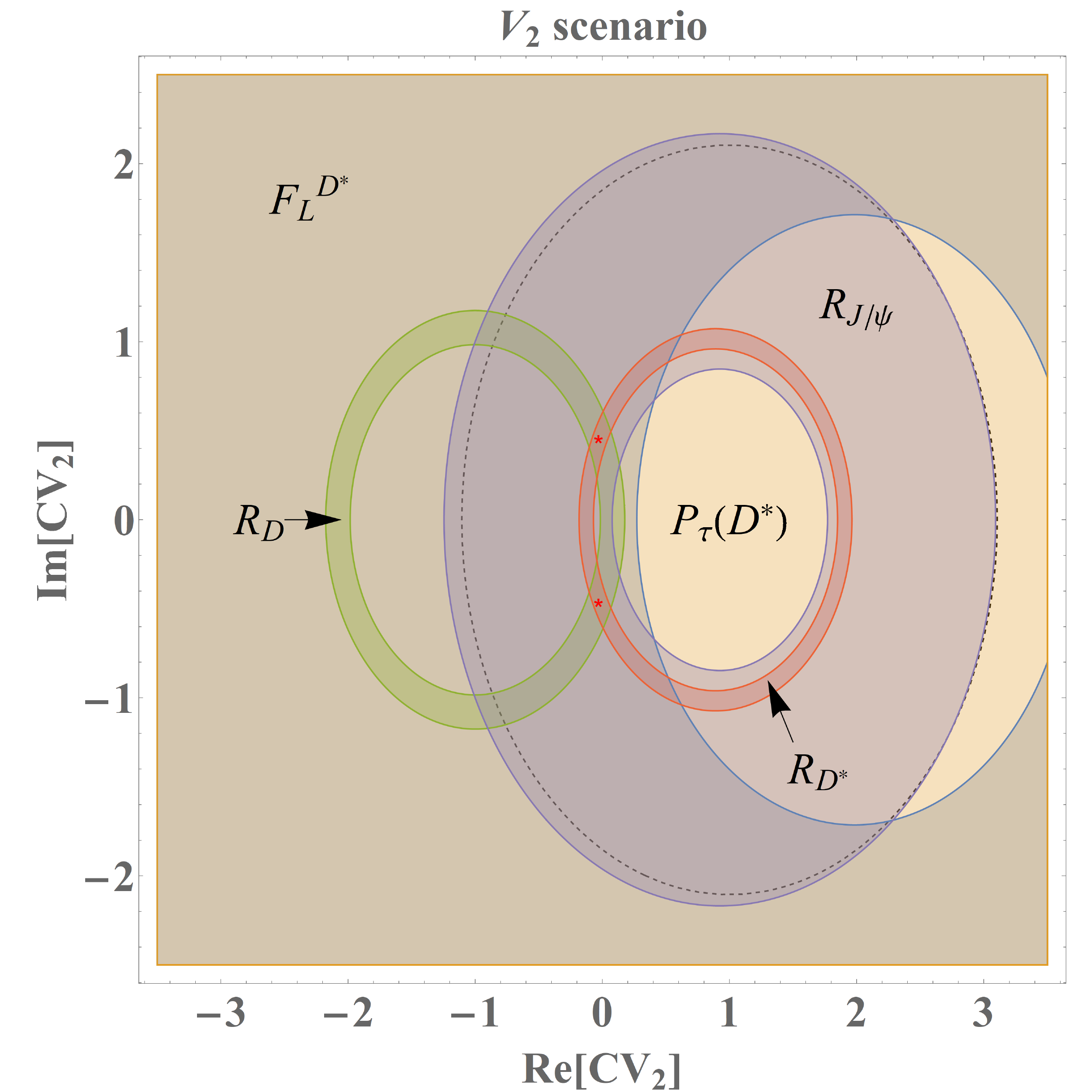}
\includegraphics[scale=0.22]{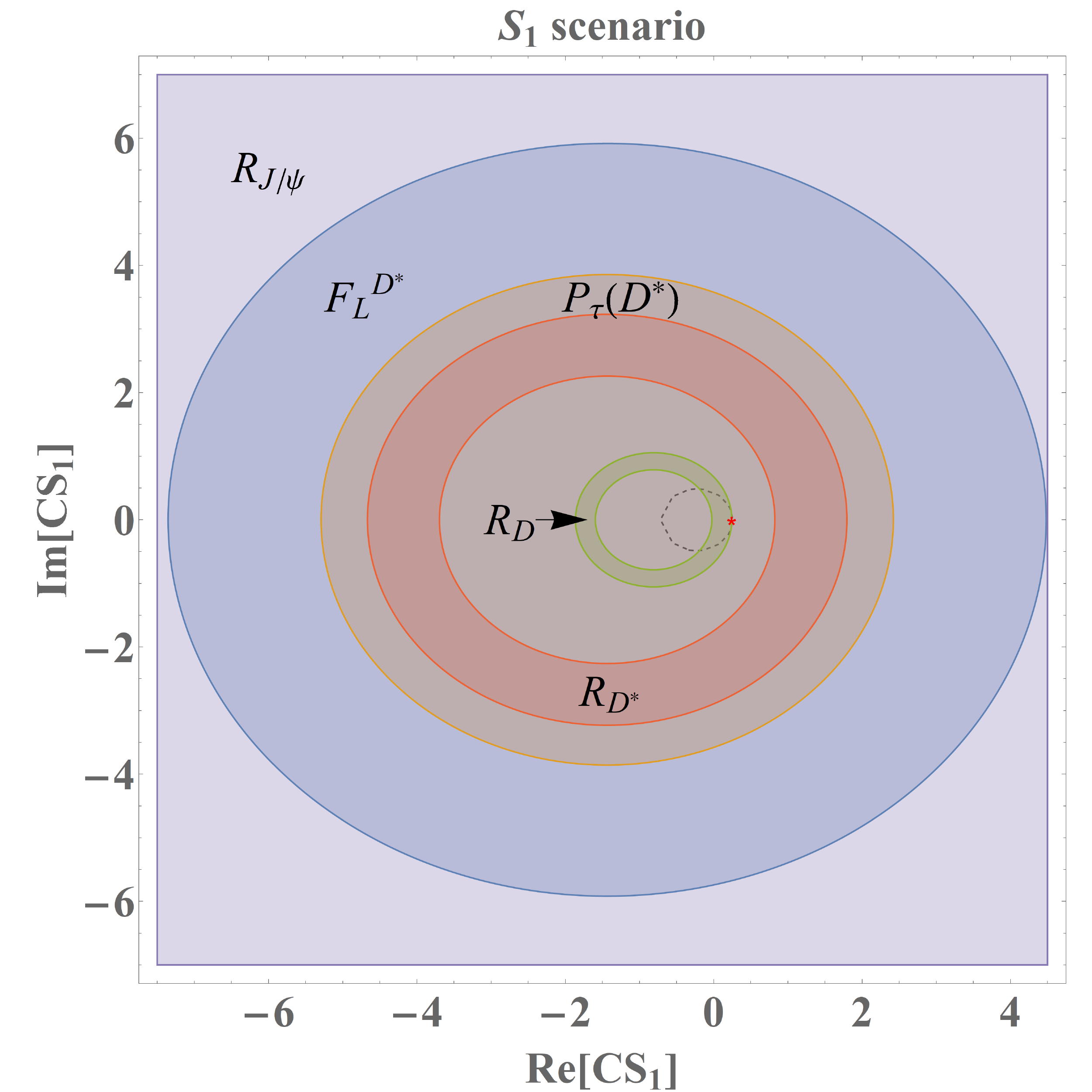}
\includegraphics[scale=0.22]{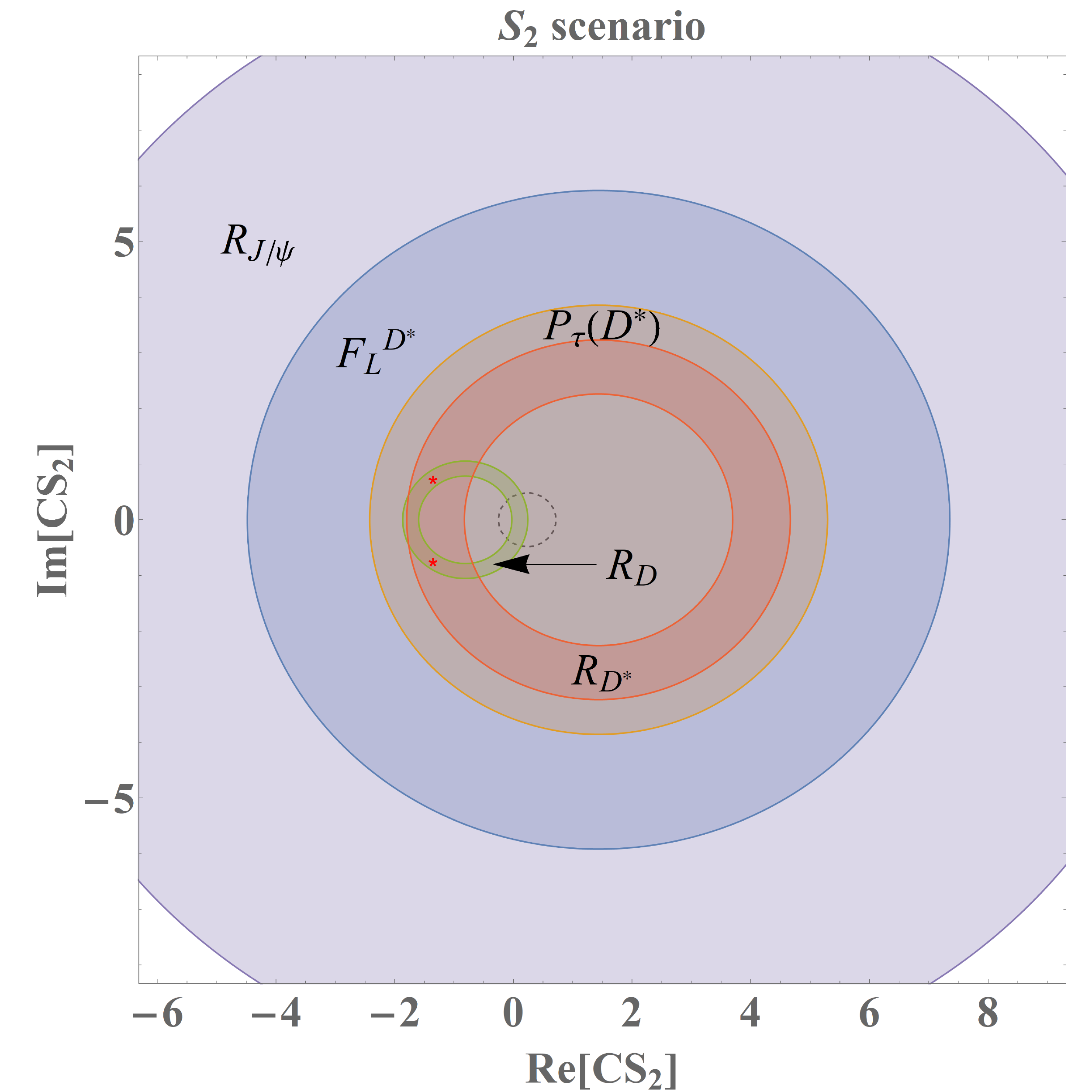}
\includegraphics[scale=0.22]{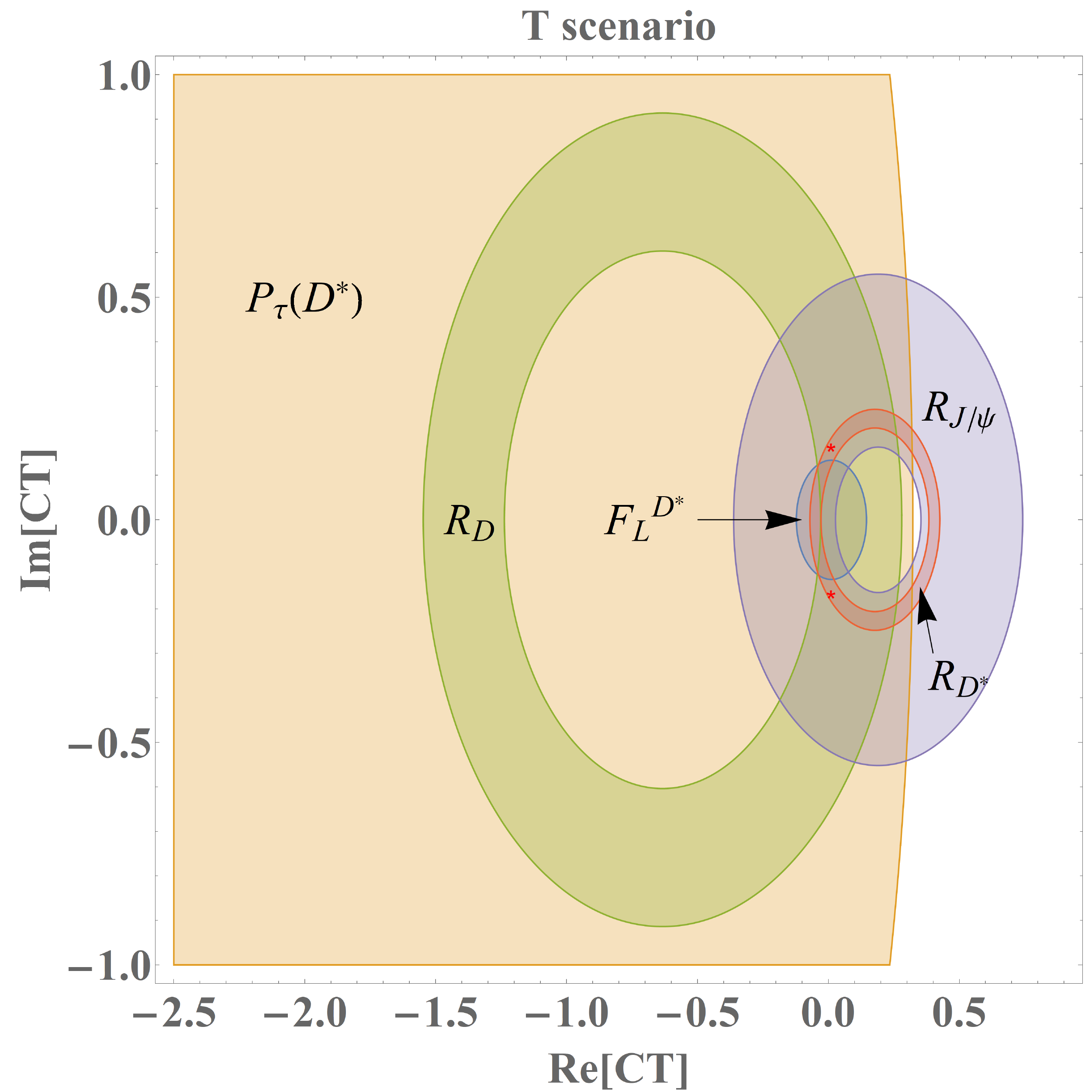}
\caption{
Constraints on the Wilson coefficients by the measurements of $R(D^{(*)})$, $R(J/\psi)$, $P_{\tau}(D^*)$ and $F_L^{D^*}$ at $95\%$ C.L. and the limit on $\mathcal B(B_c \to \tau\nu)$ (black dashed curves). The red stars and red dashed curves denote the Wilson coefficients fitted without taking into account $\mathcal B(B_c \to \tau\nu)<10\%$.}
\label{fig:constraint2}
\end{center}
\end{figure}
Assuming new physics induces simultaneously two operators with real Wilson coefficients, we also perform the global fits and the results are also listed in Table~\ref{tab:wcoef}. The $(S_1,S_2)$ scenario is excluded at $95\%$ C.L., due to the constraint $\mathcal B(B_c\to\tau\nu)<10\%$.\footnote{For this scenario, we have checked that it has no effects in decreasing the $\chi^2$ to set the Wilson coefficients to be complex, which along with the results of single scalar operator scenarios totally exclude NP models which generate purely scalar operators.}  Apart from this scenario, all the other nine scenarios have $\chi^2_{\textrm{min}}\simeq(12-15)$, which means all of them can explain the experimental data well assuming real coefficients. As shown in Figures~\ref{fig:constraint2sce1} and \ref{fig:constraint2sce2} in Appendix~\ref{app:A}, the best-fit points correspond to the parameter regions which can explain all experiments within $2\sigma$. Each of the nine scenarios has at least one natural best-fit point with a small Wilson coefficient. We also note that the scenarios including $O_{V_1}$ contain more than one best-fit point. For pure vector combination $(V_1,V_2)$ scenario, there are totally four best-fit points. Among the four points, $(0.101(21),-0.028(32))$ is natural with small Wilson coefficients, while the other three points $(-0.972(32),-1.101(21))$, $(-1.028(32),1.101(21))$, and $(-2.101(21),0.028(32))$ have $C_{V_i}\sim\mathcal{O}(1)$, which are unnatural because the SM contribution is either cancelled or flipped its sign. For each of the other three scenarios $(V_1,S_1)$, $(V_1,S_2)$, and $(V_1,T)$, there are two corresponding best-fit points. In each scenario including $O_{V_1}$, there is an unnatural best-fit point which has $1+C_{V_1}\sim-1$, with the sign of the SM contribution flipped. In the $(S_2,T)$ scenario, the global minimal $\chi^2$ point is disfavoured by the limit on $B_c\to\tau\nu$ decay, but there is another local best-fit solution which satisfies the constraint $\mathcal B(B_c\to\tau\nu)<10\%$.
\begin{table*}[!htbp]\small%
\centering
\caption{Best-fit values of the Wilson coefficients (with inclusion of $F_L^{D^*}$ and the most recent measurement of $R(D^{(*)})$ by Belle) in different NP scenarios.}
\begin{tabular}{cccc}
\hline
\hline
NP scenario  & value (without $\mathcal B(B_c\to\tau\nu)<0.1$)& $\chi^2/dof$ & Correlation \\
\hline
$V_1$       &$(1+Re[C_{V_1}])^2+(Im[C_{V_1}])^2=1.236(38)$   &$13.70/11$ & -- \\
$V_2$       &$-0.030(34)\pm0.460(52)i$   &$12.92/11$ & $\pm0.59$  \\
$S_1$       &$0.245+0.000i$   &$32.77/11$ & --  \\
$S_2$       &$-1.350(277)\pm0.740(249)i $   &$10.81/11$ & $\pm0.95$ \\
$T$       &$0.011(62)\pm0.165(60)i$   &$16.79/11$ & $\pm0.98$  \\
($V_1$, $V_2$)  & $(-0.972(32),-1.101(21))$ & $12.92/11$& 0.59 \\
($V_1$, $V_2$)  & $(-1.028(32),1.101(21))$ & $12.92/11$& 0.59\\
($V_1$, $V_2$)  & $(0.101(21),-0.028(32))$ & $12.92/11$& 0.59\\
($V_1$, $V_2$)  & $(-2.101(21),0.028(32))$ & $12.92/11$& 0.59\\
($V_1$, $S_1$) & $(0.127(28),-0.071(100))$ & $13.17/11$ & -0.78  \\
($V_1$, $S_1$) & $(-2.127(28),0.071(100))$ & $13.17/11$ & -0.78   \\
($V_1$, $S_2$) & $(0.121(20),-0.075(84))$ & $12.85/11$ & -0.51 \\
($V_1$, $S_2$) & $(-2.121(20),0.075(84))$ & $12.85/11$ & -0.51\\
($V_1$, $T$)   & $(0.087(36),-0.017(21))$ & $13.07/11$ & 0.88  \\
($V_1$, $T$)   & $(-2.087(36),0.017(21))$ & $13.07/11$ & 0.88  \\
($V_2$, $S_1$) & $(-0.119(25),0.270(54))$ & $12.13/11$ & -0.02 \\
($V_2$, $S_2$) & $(-0.158(26),0.315(63))$ & $13.19/11$ & -0.31   \\
($V_2$, $T$)   & $(0.116(57),-0.096(20))$ & $14.30/11$ & -0.90  \\
($S_1$, $S_2$) & $(0.776(97),-0.670(119))$ & $10.81/11$ & -0.83  \\
($S_1$, $T$)   & $(0.148(61),-0.050(10))$ & $13.03/11$ & 0.40 \\
($S_2$, $T$)   & $(-1.711(61),0.016(15))$ & $10.94/11$ & -0.47 \\
\hline
\hline
\end{tabular}
\label{tab:wcoef}
\end{table*}
\begin{table*}[!htbp]\small%
\centering
\caption{Best-fit values of the Wilson coefficients (with inclusion of $F_L^{D^*}$ and the most recent measurement of $R(D^{(*)})$ by Belle) in different NP scenarios.}
\begin{tabular}{cccc}
\hline
\hline
NP scenario  & value (with $\mathcal B(B_c\to\tau\nu)<0.1$)& $\chi^2/dof$ & Correlation\\
\hline
$V_1$       &$(1+Re[C_{V_1}])^2+(Im[C_{V_1}])^2=1.236(38)$   &$13.70/11$ & -- \\
$V_2$       &$-0.030(34)\pm0.460(52)i$   &$12.92/11$ & $\pm0.59$  \\
$S_1$       &$0.245+0.000i$   &$32.77/11$ & --  \\
$S_2$       &$0.072\pm0.461i $   &$39.10/11$ & -- \\
$T$       &$0.011(62)\pm0.165(60)i$   &$16.79/11$ & $\pm0.98$  \\
($V_1$, $V_2$)  & $(-0.972(32),-1.101(21))$ & $12.92/11$& 0.59 \\
($V_1$, $V_2$)  & $(-1.028(32),1.101(21))$ & $12.92/11$& 0.59\\
($V_1$, $V_2$)  & $(0.101(21),-0.028(32))$ & $12.92/11$& 0.59\\
($V_1$, $V_2$)  & $(-2.101(21),0.028(32))$ & $12.92/11$& 0.59\\
($V_1$, $S_1$) & $(0.127(28),-0.071(100))$ & $13.17/11$ & -0.78  \\
($V_1$, $S_1$) & $(-2.127(28),0.071(100))$ & $13.17/11$ & -0.78   \\
($V_1$, $S_2$) & $(0.121(20),-0.075(84))$ & $12.85/11$ & -0.51 \\
($V_1$, $S_2$) & $(-2.121(20),0.075(84))$ & $12.85/11$ & -0.51\\
($V_1$, $T$)   & $(0.087(36),-0.017(21))$ & $13.07/11$ & 0.88  \\
($V_1$, $T$)   & $(-2.087(36),0.017(21))$ & $13.07/11$ & 0.88  \\
($V_2$, $S_1$) & $(-0.117(25),0.229(57))$ & $12.72/11$ & 0.01 \\
($V_2$, $S_2$) & $(-0.158(26),0.315(63))$ & $13.19/11$ & -0.31   \\
($V_2$, $T$)   & $(0.116(57),-0.096(20))$ & $14.30/11$ & -0.90  \\
($S_1$, $S_2$) & $(-0.785,-1.041)$ & $32.01/11$ & --  \\
($S_1$, $T$)   & $(0.148(61),-0.050(10))$ & $13.03/11$ & 0.40 \\
($S_2$, $T$)   & $(0.143(63),-0.057(9))$ & $13.72/11$ & 0.14 \\
\hline
\hline
\end{tabular}
\label{tab:wcoef2}
\end{table*}

Furthermore, we give the predictions in Table~\ref{tab:obser1} and Table~\ref{tab:obser12} for various observables in the SM and the allowed model-independent scenarios. For the SM values, we observe $\sim4\sigma$ deviation from the experimental averages, which agrees with the recent findings in \cite{Iguro:2020cpg} and \cite{Bordone:talk}. Compared with the recent results obtained using the HQET parametrizations, our prediction for $R(D^{*})$ is consistent with the optimal result (2/1/0 fit) in \cite{Iguro:2020cpg} and the theoretical prediction (without including experimental information) in \cite{Bordone:2019vic} within $1\sigma$, and our prediction for $R(D)$ agrees with \cite{Iguro:2020cpg} within $1\sigma$ and \cite{Bordone:2019vic} within $1.6\sigma$\footnote{The slight variation has two major sources: parametrization dependence and different use of input data in the fit. In \cite{Bordone:2019vic}, full $1/m_c^2$ corrections and higher-order z expansion of Isgur wise functions were included in the HQET parametrization with more fitting parameters. The slight parametrization dependence of $R(D^{(*)})$ can also be reflected in \cite{Iguro:2020cpg} by comparing the results obtained in the 3/2/1 and the 2/1/0 fits. On the other hand, we have not considered as in \cite{Bordone:2019vic} the QCD sum rule (QCDSR) results for sub-leading Isgur-Wise functions~\cite{Neubert:1992wq,Neubert:1992pn,Ligeti:1993hw} in our fit, which can lead to $1\sigma$ difference in certain scenarios as shown in \cite{Bernlochner:2017jka}.}.

In contrast to the SM, each of these NP scenarios can well explain the data of $R(D^{(*)})$, but the polarization and angular observables $P_{\tau}(D)$, $P_{\tau}(D^*)$, $F_L^{D^*}$, $A_{\textrm{FB}}(D)$, and $A_{\textrm{FB}}(D^*)$ will be useful to distinguish different scenarios at ongoing or future experiments like Belle II. Among those quantities, $P_{\tau}(D^*)$ and $F_L^{D^*}$ have already been measured at Belle \cite{Hirose:2016wfn,Adamczyk:2019wyt,Abdesselam:2019wbt} as listed in Table~\ref{tab:exdata}, while $P_{\tau}(D)$ and $A_{\textrm{FB}}$ have not been measured. In most scenarios, the predictions for $P_{\tau}(D)$ are close to that in the SM. But in the scenarios $(V_2,S_1)$, $(V_2,S_2)$, $(S_1,T)$, and $(S_2,T)$, we predict larger $P_{\tau}(D)$. Similarly, in most scenarios, the predictions on $P_{\tau}(D^*)$ are also close to the SM prediction, but $T$, $(V_2,S_1)$, $(V_2,T)$, $(S_1,T)$ scenarios predict larger $P_{\tau}(D^*)$ than the SM, while $(V_2,S_2)$ scenario predicts smaller $P_{\tau}(D^*)$. Precision measurements on $\tau$ polarization will be helpful to distinguish among different scenarios. For the $D^*$ polarization, predictions by the SM and most scenarios are close to $0.47$, but in $T$, $(V_2,T)$ and $(S_2,T)$ scenarios the predictions are smaller, while in $(V_2,S_1)$ scenario it is bigger; the pure tensor scenario also predicts small $F_L^{D^*}$ but with large uncertainty. In all scenarios, we cannot predict a $F_L^{D^*}$ larger than $0.5$, because the scalar contributions are suppressed by the $B_c\to\tau\nu$ decay constraint. In the SM and most NP scenarios, $A_{\textrm{FB}}(D)$ is predicted to be close to $0.36$, except that in the last five scenarios in Table~\ref{tab:obser12} without $O_{V_1}$, smaller $A_{\textrm{FB}}(D)$($\simeq0.31-0.33$) are predicted. Therefore the measurement of $A_{\textrm{FB}}(D)$ at future experiments may be useful to distinguish between these two classes of NP scenarios. The SM and most of the new physics scenarios predict $A_{\textrm{FB}}(D^*)\simeq-0.05$, while other scenarios predict $A_{\textrm{FB}}(D^*)$ closer to zero. The most special solutions are two of the $(V_1,V_2)$ solutions, corresponding to which the SM contributions are almost canceled by the NP contributions and the $V_2$ contributions dominate the $b\to c\tau\nu$ decay. Such solutions predict $A_{\textrm{FB}}(D^*)=0.318(5)(9)$ which is significantly larger than all other cases. It can be easily distinguished from other cases in experiments.
\begin{table*}[!htbp]\small%
\centering
\caption{Predictions for $R(D)$, $R(D^*)$, $P_\tau(D)$ and $P_\tau(D^*)$ by the SM and the model-independent NP scenarios. The first and second uncertainties are respectively due to the input parameters and the fitted Wilson coefficients.}
\begin{tabular}{ccccc}
\hline
\hline
Scenario & $R(D)$& $R(D^*)$   &$P_\tau(D)$ & $P_\tau(D^*)$ \\
\hline
SM             &$0.289(5)(0)$ & $0.237(8)(0)$ &$0.328(3)(0)$ &$-0.490(5)(0)$ \\
$V_1$       &$0.358(6)(11)$ & $0.293(10)(9)$ &$0.328(3)(0)$ &$-0.490(5)(0)$ \\
$V_2$       &$0.334(6)(30)$ & $0.300(10)(12)$ &$0.328(3)(0)$ &$-0.490(5)(1)$ \\
$T$       &$0.300(5)(26)$ & $0.303(21)(34)$ &$0.314(3)(48)$ &$-0.357(25)(74)$ \\
$(V_1,V_2)$ &$0.333(6)(31)$ & $0.300(10)(13)$ & $0.328(3)(0)$ & -0.490(5)(1) \\
$(V_1,S_1)$ &$0.337(6)(30)$ & $0.298(10)(13)$ & $0.268(3)(88)$ & -0.502(4)(16) \\
$(V_1,S_2)$ &$0.332(5)(30)$ & $0.300(10)(12)$ & $0.264(3)(75)$ & -0.478(5)(14) \\
$(V_1,T)$ &$0.336(6)(30)$ & $0.299(10)(15)$ & $0.340(3)(15)$ & -0.479(4)(17) \\
$(V_2,S_1)$ &$0.318(6)(31)$ & $0.297(10)(13)$ & $0.523(3)(40)$ & $-0.447(7)(10)$ \\
$(V_2,S_2)$ &$0.333(6)(31)$ & $0.299(10)(12)$ & $0.587(3)(42)$ & $-0.535(3)(9)$ \\
$(V_2,T)$ &$0.328(6)(28)$ & $0.299(21)(12)$ & $0.396(2)(12)$ & $-0.402(12)(23)$ \\
$(S_1,T)$ &$0.337(6)(30)$ & $0.299(13)(12)$ & $0.486(3)(41)$ & $-0.428(5)(9)$ \\
$(S_2,T)$ &$0.333(6)(29)$ & $0.300(15)(12)$ & $0.487(3)(45)$ & $-0.463(7)(13)$ \\
\hline
\hline
\end{tabular}
\label{tab:obser1}
\end{table*}
\begin{table*}[!htbp]\small%
\centering
\caption{Predictions for $F_L^{D^*}$, $\mathcal A_{FB}(D)$ and $A_{FB}(D^*)$ by the SM and the model-independent NP scenarios. The first and second uncertainties are respectively due to the input parameters and the fitted Wilson coefficients.}
\begin{tabular}{cccc}
\hline
\hline
Scenario  &$F_L^{D^*}$ &$\mathcal A_{FB}(D)$& $A_{FB}(D^*)$\\
\hline
SM          &$0.467(4)(0)$ & 0.360(1)(0) & -0.057(6)(0)\\
$V_1$       &$0.467(4)(0)$ & 0.360(1)(0) & -0.057(6)(0)\\
$V_2$        &$0.470(4)(3)$ & 0.360(1)(0) & 0.016(4)(10)\\
$T$      &$0.401(13)(39)$ & 0.357(1)(25) & 0.013(15)(20)\\
$(V_1,V_2)$  & $0.470(4)(3)$ & 0.360(1)(0)& 0.318(5)(9)/-0.047(4)(11)\\
$(V_1,S_1)$  & $0.463(4)(6)$ & 0.365(1)(7)& -0.063(4)(9)\\
$(V_1,S_2)$  & $0.472(4)(5)$ & 0.365(1)(5)& -0.050(4)(7)\\
$(V_1,T)$  & $0.463(4)(7)$ & 0.352(1)(10)& -0.039(4)(25)\\
$(V_2,S_1)$  & $0.491(4)(4)$ & 0.327(2)(9)& 0.003(3)(9)\\
$(V_2,S_2)$  &$0.463(4)(3)$ & 0.311(2)(12)& -0.031(4)(9)\\
$(V_2,T)$ & $0.423(7)(12)$ & 0.310(2)(9)& 0.011(9)(9)\\
$(S_1,T)$  & $0.459(5)(7)$ & 0.313(2)(8)& 0.012(6)(9)\\
$(S_2,T)$ & $0.440(5)(4)$ & 0.309(2)(10)& -0.007(7)(12)\\
\hline
\hline
\end{tabular}
\label{tab:obser12}
\end{table*}
\newpage
\subsection{Leptoquark models}
\begin{figure}[!htbp]
\begin{center}
\includegraphics[scale=0.22]{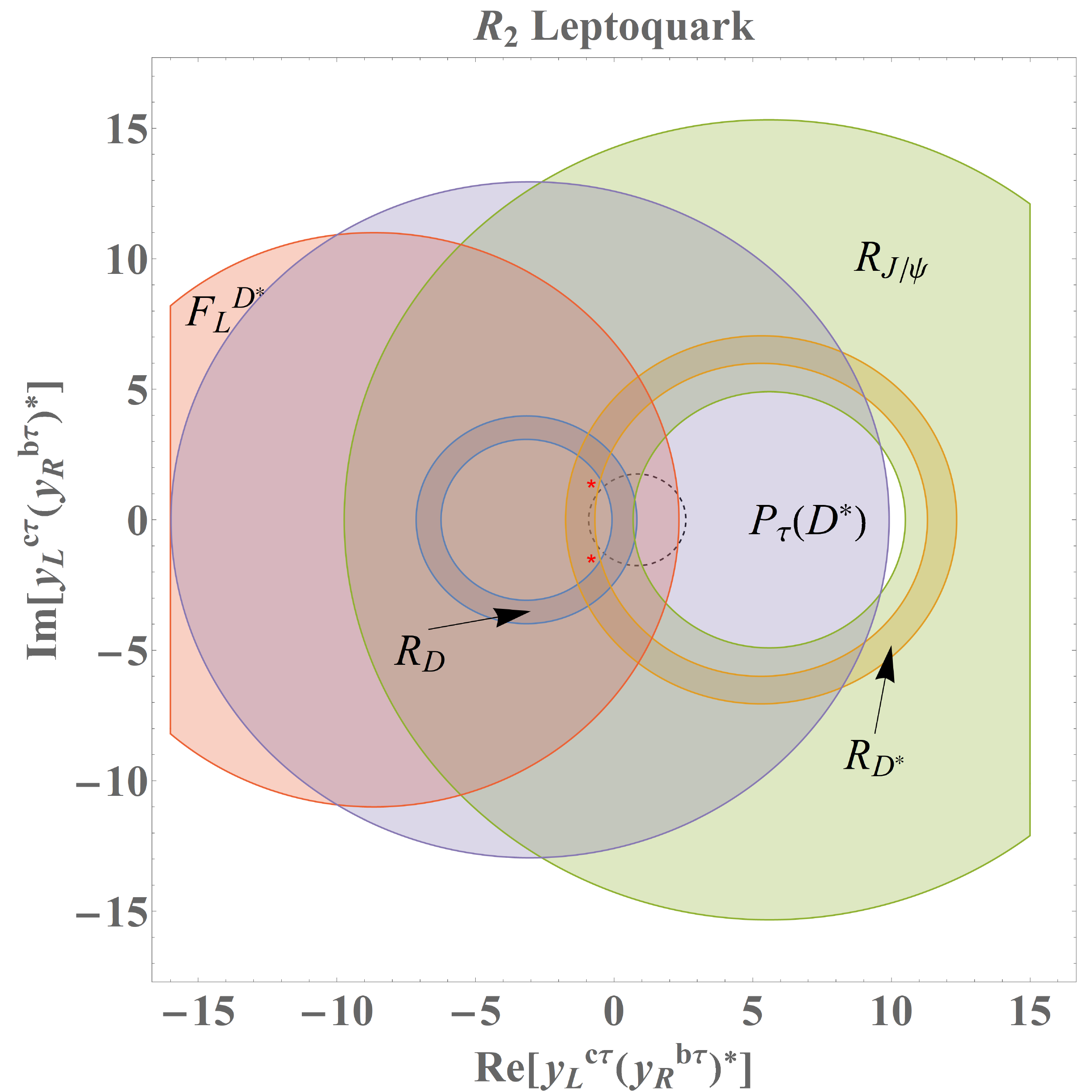}
\includegraphics[scale=0.22]{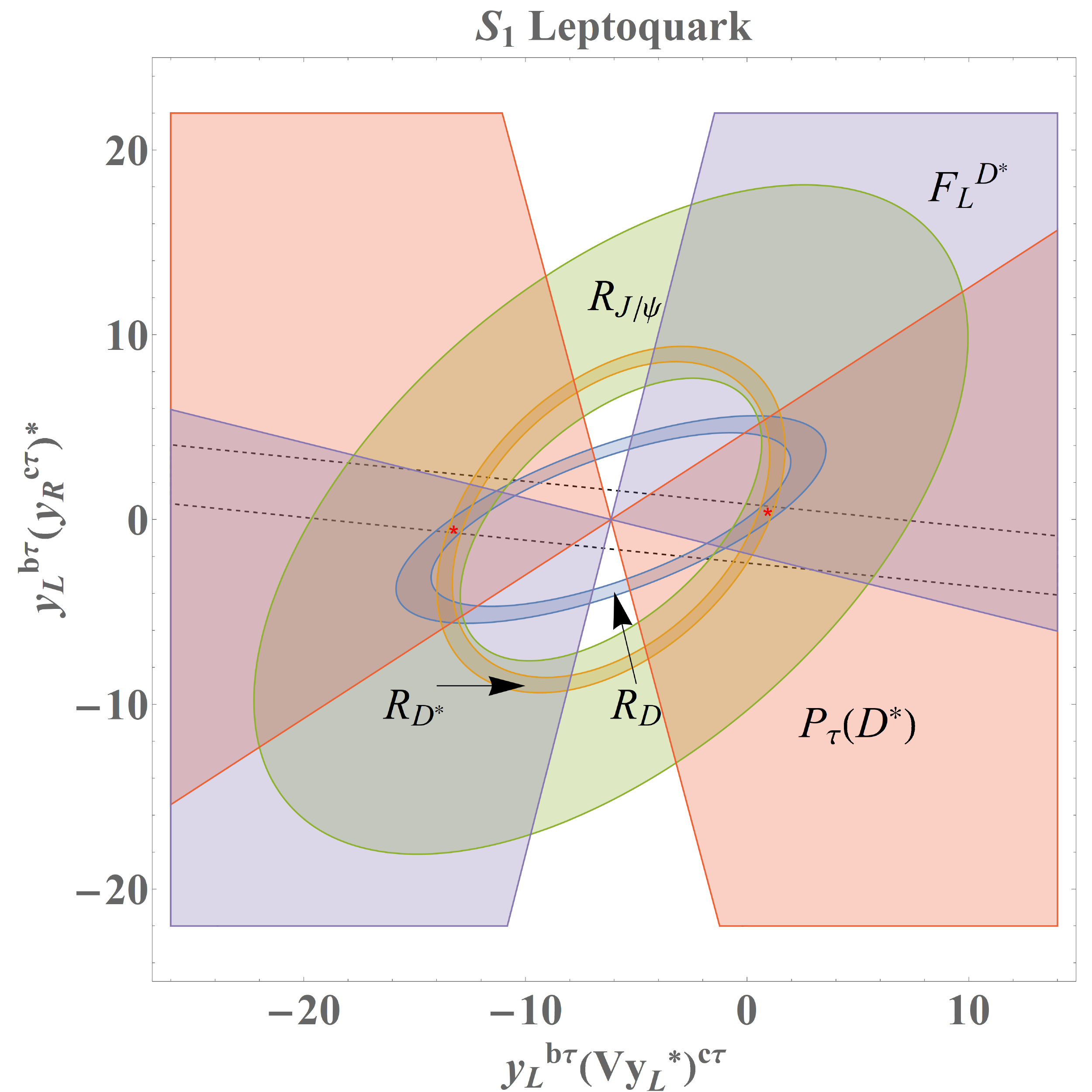}
\includegraphics[scale=0.22]{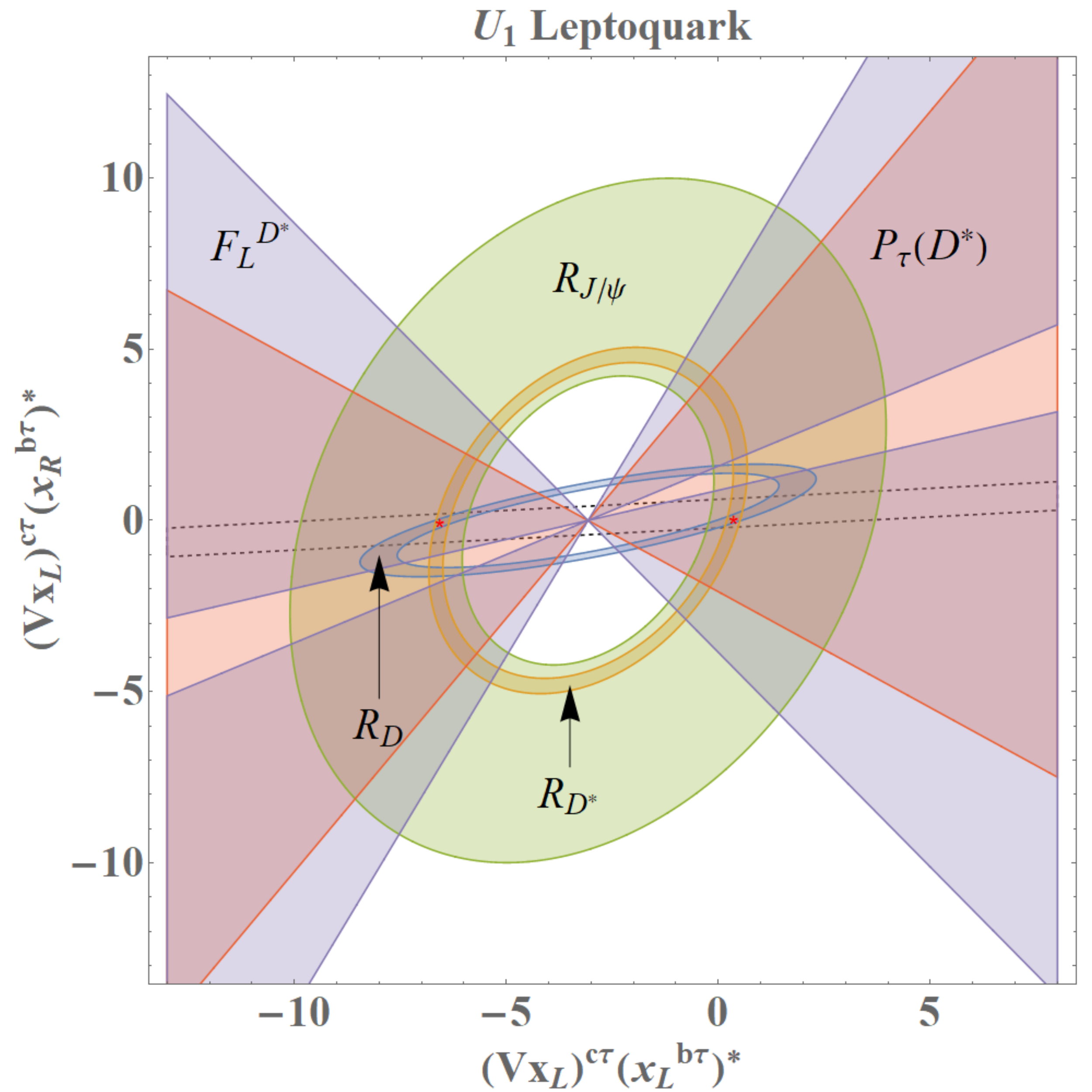}
\caption{Constraints on the leptoquark couplings by the measurements of $R(D^{(*)})$, $R(J/\psi)$, $P_{\tau}(D^*)$ and $F_L^{D^*}$ at $95\%$ C.L. and the limit on $\mathcal B(B_c \to \tau\nu)$ (black dashed curves). The red stars and red dashed curves denote the Wilson coefficients fitted without taking into account $\mathcal B(B_c \to \tau\nu)<10\%$.}
\label{fig:constraint2LQ}
\end{center}
\end{figure}

\begin{figure}[!htbp]
\begin{center}
\includegraphics[scale=0.22]{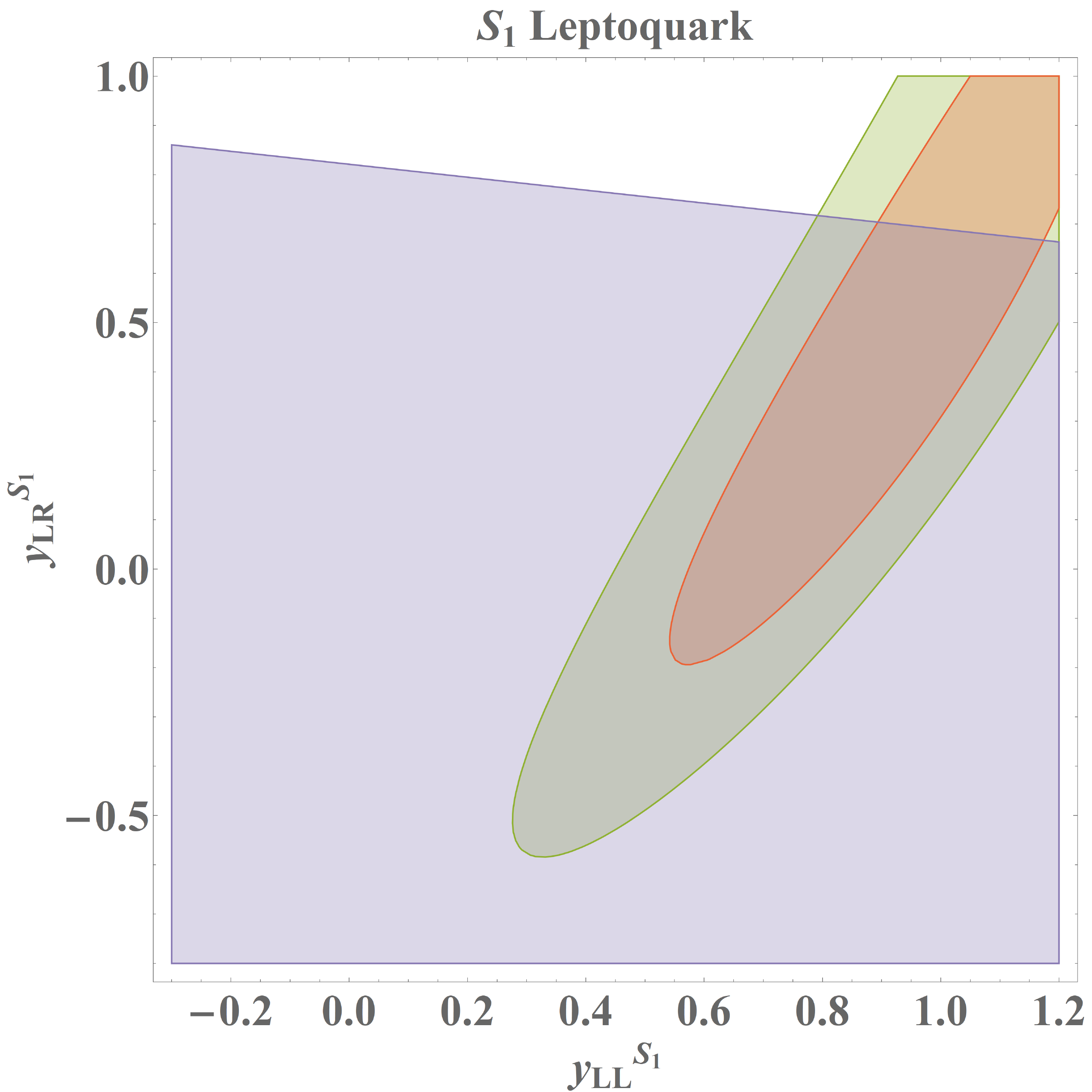}
\includegraphics[scale=0.22]{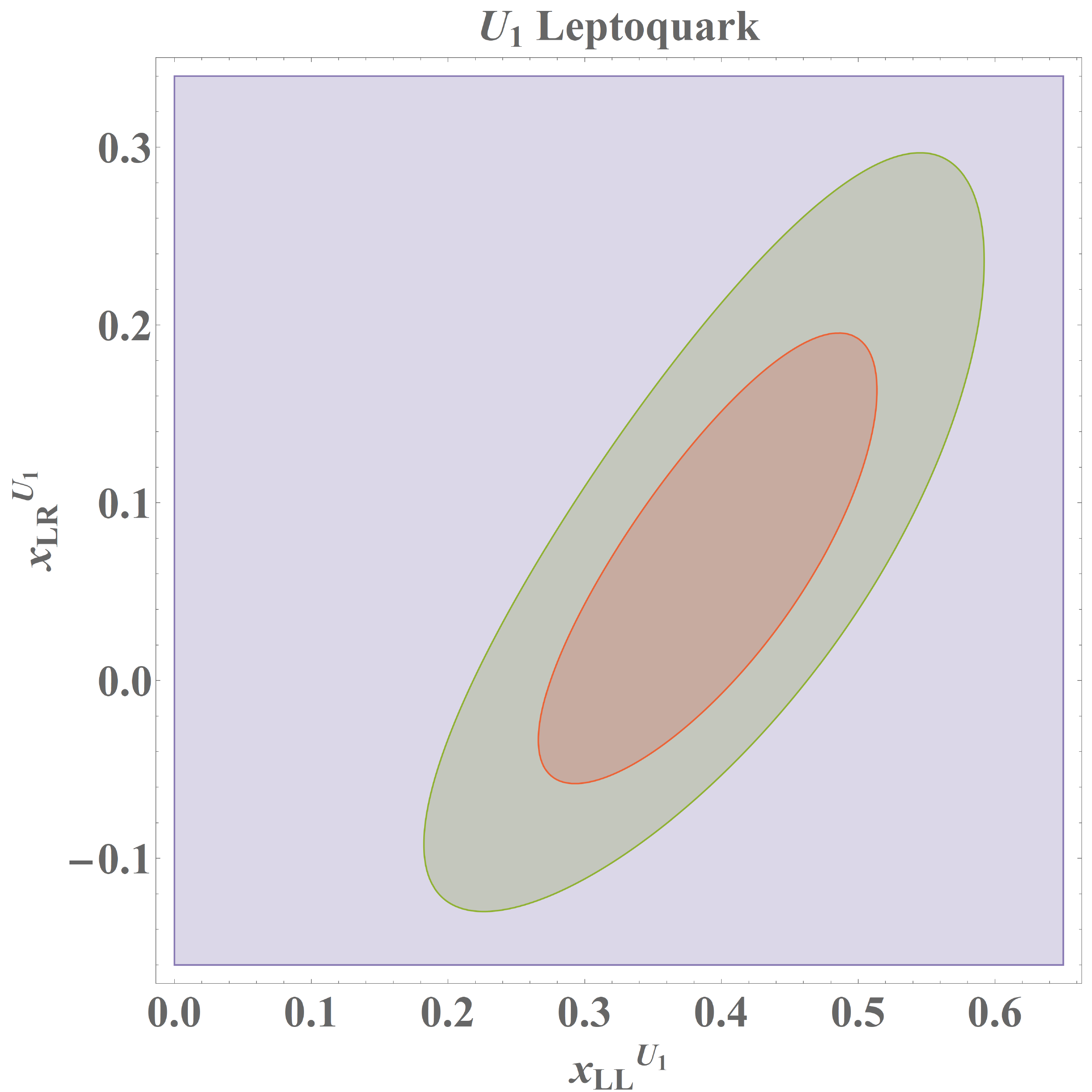}
\includegraphics[scale=0.22]{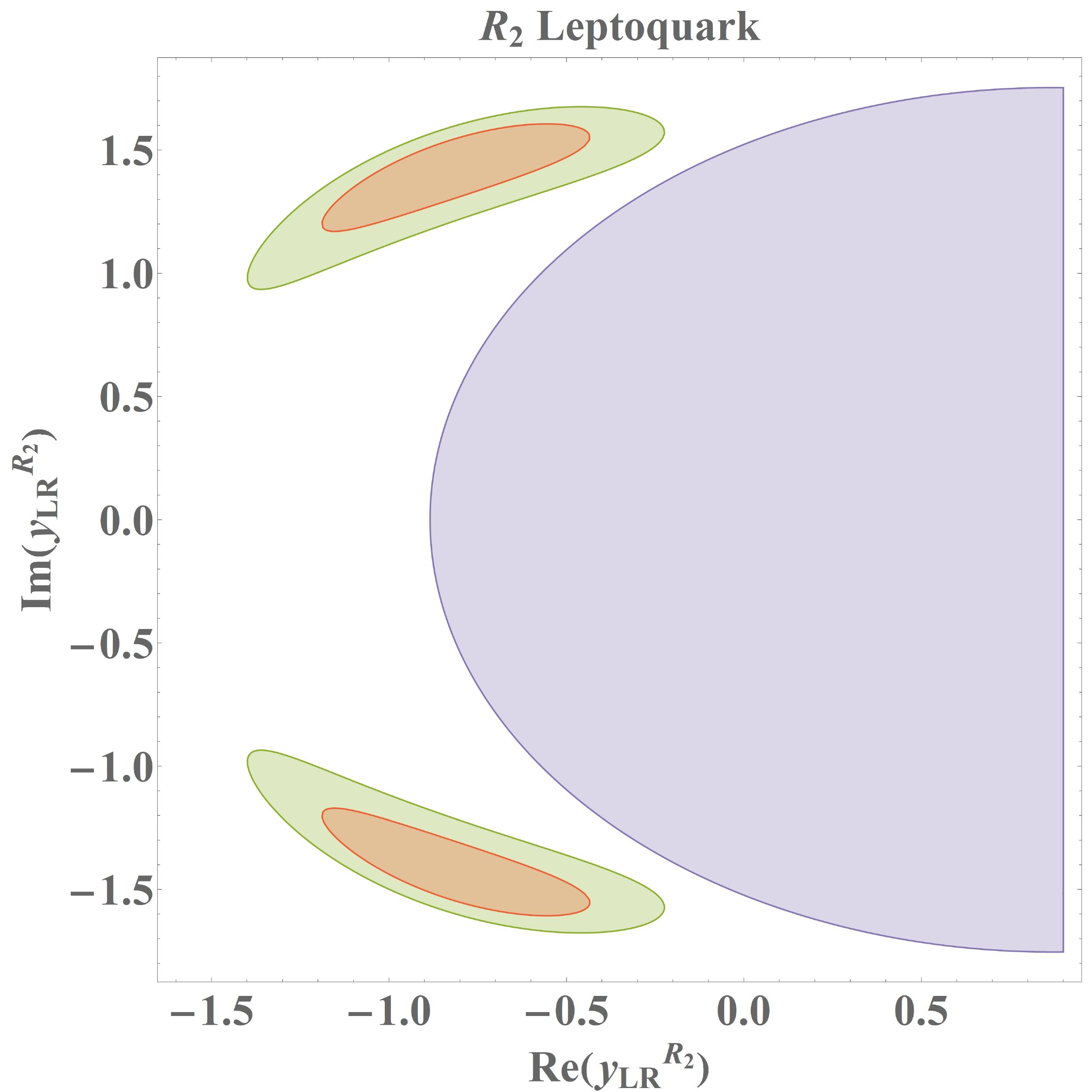}
\caption{
Allowed regions for LQ couplings in $S_1$ (upper left), $U_1$ (upper right), and $R_2$ (lower) LQ models. The light red regions are allowed at $68\%$ C.L. and the light green regions are allowed at $95\%$ C.L. without the $\mathcal{B}(B_c\rightarrow\tau\nu)$ constraint. In the light purple regions, $\mathcal{B}(B_c\rightarrow\tau\nu)\leq0.1$, and the dashed lines are the boundaries of such regions. We ignore the unnatural best-fit points with large couplings (or Wilson coefficients) for $S_1$ and $U_1$ models. We can see clearly that in the $R_2$ LQ model, the allowed regions locate outside the region with $\mathcal{B}(B_c\rightarrow\tau\nu)\leq0.1$.}
\label{fig:LQregion}
\end{center}
\end{figure}

If we do not discuss the exact ultraviolet (UV) completions and consider only the standard model fermions (without right-handed neutrinos), there can be ten types of LQs \cite{Tanabashi:2018oca,Dorsner:2016wpm,Buchmuller:1986zs}, three of which, namely $R_2$, $S_1$, and $U_1$ LQs with quantum numbers listed in Table~\ref{tab:lq} are possible explanations of the $b\to c\tau\nu$ anomalies as discussed in \cite{Sakaki:2013bfa,Angelescu:2018tyl,Dorsner:2016wpm,Iguro:2018vqb}.
\begin{table}[!htbp]\small%
\centering
\caption{Quantum numbers and couplings of the leptoquarks which can explain the $b\rightarrow c\tau\nu$ anomalies. Here $F\equiv3B+L$.}
\begin{tabular}{ccccc}
\hline
\hline
&\begin{tabular}{c}SM quantum number\\$[\textrm{SU}(3)\times\textrm{SU}(2)\times\textrm{U}(1)]$\end{tabular}&$F$&Spin&Fermions coupled to\\
\hline
$R_2$&$(3,2,7/6)$&$0$&$0$&$\bar{c}_R\nu_L,\bar{b}_L\tau_R$\\
$S_1$&$(\bar{3},1,1/3)$&$-2$&$0$&$\bar{b}_L^c\nu_L,\bar{c}_L^c\tau_L,\bar{c}_R^c\tau_R$\\
$U_1$&$(3,1,2/3)$&$0$&$1$&$\bar{c}_L\gamma_{\mu}\nu_L,\bar{b}_L\gamma_{\mu}\tau_L,\bar{b}_R\gamma_{\mu}\tau_R$\\
\hline
\hline
\end{tabular}
\label{tab:lq}
\end{table}
The corresponding interactions are then \cite{Angelescu:2018tyl,Iguro:2018vqb}
\begin{align}
\mathcal{L}_{R_2}=\left(y_R^{b\tau}\bar{b}_L\tau_R+y_L^{c\tau}\bar{c}_R\nu_L\right)Y_{2/3}+\textrm{H.c.};\\
\mathcal{L}_{S_1}=\left((V^*_{\textrm{CKM}}y_L)^{c\tau}\bar{c}^c_L\tau_L-y_L^{b\tau}\bar{b}^c_L\nu_L+y_R^{c\tau}\bar{c}^c_R\tau_R\right)Y_{1/3}+\textrm{H.c.};\\
\mathcal{L}_{U_1}=\left((V_{\textrm{CKM}}x_L)^{c\tau}\bar{c}_L\gamma_{\mu}\nu_L+x_L^{b\tau}\bar{b}_L\gamma_{\mu}\tau_L+x_R^{b\tau}\bar{b}_R\gamma_{\mu}\tau_R\right)X^{\mu}_{2/3}+\textrm{H.c.};
\end{align}
where $f^c$ denotes the charged conjugation of a fermion $f$, and $X(Y)_Q$ are the are vector (scalar) LQs with the electric charge $Q$. Integrating out the heavy LQ at the scale $m_{\textrm{LQ}}$, the Wilson coefficients are obtained~\cite{Iguro:2018vqb}:
\begin{align}
C_{S_2}(m_{\textrm{LQ}})=4C_T(m_{\textrm{LQ}})=\frac{y_L^{c\tau}(y_R^{b\tau})^*}{4\sqrt{2}G_FV_{cb}m^2_{\textrm{LQ}}},~~~~~&(R_2~\textrm{LQ});
\end{align}
\begin{align}
C_{V_1}(m_{\textrm{LQ}})=\frac{y_L^{b\tau}(V_{\textrm{CKM}}y_L^*)^{c\tau}}{4\sqrt{2}G_FV_{cb}m^2_{\textrm{LQ}}},~ C_{S_2}(m_{\textrm{LQ}})=-4C_T(m_{\textrm{LQ}})=-\frac{y_L^{b\tau}(y_R^{c\tau})^*}{4\sqrt{2}G_FV_{cb}m^2_{\textrm{LQ}}},~&(S_1~\textrm{LQ});
\end{align}
\begin{align}
C_{V_1}(m_{\textrm{LQ}})=\frac{(V_{\textrm{CKM}}x_L)^{c\tau}(x_L^{b\tau})^*}{2\sqrt{2}G_FV^2_{cb}m^2_{\textrm{LQ}}},~~ C_{S_1}(m_{\textrm{LQ}})=-\frac{(V_{\textrm{CKM}}x_L)^{c\tau}(x_R^{b\tau})^*}{\sqrt{2}G_FV^2_{cb}m^2_{\textrm{LQ}}},~~&(U_1~\textrm{LQ}).
\end{align}
Following the notations in \cite{Iguro:2018vqb}, we denote
\begin{align}
&y_{LR}^{R_2}\equiv y_L^{c\tau}(y_R^{b\tau})^*,\quad y_{LL}^{S_1}\equiv y_L^{b\tau}(V_{\textrm{CKM}}y_L^*)^{c\tau},\nonumber\\
&y_{LR}^{S_1}\equiv y_L^{b\tau}(y_R^{c\tau})^*, \quad x_{LL(LR)}^{U_1}\equiv (V_{\textrm{CKM}}x_L)^{c\tau}(x_{L(R)}^{b\tau})^*.
\end{align}
We choose $m_{\textrm{LQ}}=1.5~\textrm{TeV}$ as the benchmark in our numerical analysis, since such a mass for LQ is still allowed at the LHC~\cite{Tanabashi:2018oca}. After performing the three-loop QCD and one-loop EW running, the coefficients at the low scale $m_b$ are then \cite{Iguro:2018vqb,Gonzalez-Alonso:2017iyc}
\begin{align}
&\left(\begin{array}{c}C_{S_1}(m_b)\\C_{S_2}(m_b)\\C_T(m_b)\end{array}\right)=
\left(\begin{array}{ccc}1.788&&\\&1.789&-0.340\\&-4.43\times10^{-3}&0.837\end{array}\right)
\left(\begin{array}{c}C_{S_1}(m_{\textrm{LQ}})\\C_{S_2}(m_{\textrm{LQ}})\\C_T(m_{\textrm{LQ}})\end{array}\right);&\\
&C_{V_1}(m_b)=C_{V_1}(m_{\textrm{LQ}}).
\end{align}
For $R_2$ LQ model, we allow the single coupling to be complex, while for $S_1$ and $U_1$ LQ models, we fix the couplings $y_{LL(LR)}^{S_1}$ and $x_{LL(LR)}^{U_1}$ to be real. We find $\chi^2_{\textrm{min}}\approx13$ for both $S_1$ and $U_1$ models, and the best-fit points without imposing the constraint from the $B_c\rightarrow\tau\nu$ decay automatically satisfies $\mathcal B(B_c\rightarrow\tau\nu)<10\%$ as shown in Table~\ref{tab:wcoef3}, Table \ref{tab:wcoef1} and Figure~\ref{fig:constraint2LQ}. Both models can generate the operator $O_{V_1}$, and in each model, similar to the model-independent scenarios involving $\mathcal O_{V_1}$, there is another best-fit point with $1+C_{V_1}\sim-1$ , which is unnatural. For $R_2$ LQ model, the best-fit result doesn't accommodate $\mathcal B(B_c\rightarrow\tau\nu)<10\%$, and by directly imposing the constraint we obtain $\chi^2_{\textrm{min}}\approx22.82>19.7$, which means it is not allowed at $95\%$ C.L. We show the $68\%$ C.L. (green) and $95\%$ C.L. (yellow) allowed regions of the LQ couplings for all the three LQ models in Figure~\ref{fig:LQregion}.
\begin{table*}[!htbp]\small%
\centering
\caption{Best-fit values of the Wilson coefficients (with inclusion of $F_L^{D^*}$ and the most recent measurement of $R(D^{(*)})$ by Belle) in leptoquark models.}
\begin{tabular}{cccc}
\hline
\hline
LQ Type  & value (without $\mathcal B(B_c\to\tau\nu)<0.1$)& $\chi^2/dof$ & corr \\
\hline
$R_2$       &$(Re[y_L^{c\tau}(y_R^{b\tau})^*],Im[y_L^{c\tau}(y_R^{b\tau})^*])=(-0.811(234),\pm1.439(128))$   &12.87/11 &$\pm0.82$  \\
$S_1$       &$(y_L^{b\tau}(Vy_L^*)^{c\tau}),y_L^{b\tau}(y_R^{c\tau})^*)=(0.936(270),0.476(509))$   &$12.70/11$  & 0.92\\
$S_1$       &$(y_L^{b\tau}(Vy_L^*)^{c\tau}),y_L^{b\tau}(y_R^{c\tau})^*)=(-13.224(270),-0.476(509))$   &$12.70/11$  & 0.92\\
$U_1$      &$((Vx_L)^{c\tau}(x_L^{b\tau})^*,(Vx_L)^{c\tau}(x_R^{b\tau})^*)=(0.391(85),0.061(86))$  &$13.17/11$ & 0.78 \\
$U_1$      &$((Vx_L)^{c\tau}(x_L^{b\tau})^*,(Vx_L)^{c\tau}(x_R^{b\tau})^*)=(-6.536(85),-0.061(86))$  &$13.17/11$ & 0.78  \\
\hline
\hline
\end{tabular}
\label{tab:wcoef3}
\end{table*}
\begin{table*}[!htbp]\small%
\centering
\caption{Best-fit values of the Wilson coefficients (with inclusion of $F_L^{D^*}$ and the most recent measurement of $R(D^{(*)})$ by Belle) in leptoquark models.}
\begin{tabular}{cccc}
\hline
\hline
LQ Type   & value (with $\mathcal B(B_c\to\tau\nu)<0.1$)& $\chi^2/dof$ & corr \\
\hline
$R_2$      &$(-0.164(398),\pm1.446(117))$ &22.87/11& $\pm0.29$ \\
$S_1$     &$(0.936(270),0.476(509))$ &$12.70/11$  & 0.92\\
$S_1$     &$(-13.224(270),-0.476(509))$ &$12.70/11$  & 0.92\\
$U_1$      &$(0.391(85),0.061(86))$  &$13.17/11$  & 0.78\\
$U_1$      &$(-6.535(85),-0.061(86))$   &$13.17/11$ & 0.78\\
\hline
\hline
\end{tabular}
\label{tab:wcoef1}
\end{table*}

For $m_{\textrm{LQ}}=1.5~\textrm{TeV}$, both $S_1$ and $U_1$ LQ models with relatively small couplings can explain the anomalies and satisfy other experimental constraints at the same time. In these two models, fits for real couplings are good which means that CP-conserving scenarios are allowed. Such results also hold for heavier LQs, but the exact allowed region will change slightly depending on the masses of LQs. For $R_2$ LQ model, to explain the $R(D^{(*)})$ anomalies, there must be large CP-violation phase in the LQ couplings. But unfortunately, we can see clearly from the lower pannel in Figure~\ref{fig:LQregion} that, the allowed regions without $\mathcal{B}(B_c\rightarrow\tau\nu)$ constraint locate outside the region with $\mathcal{B}(B_c\rightarrow\tau\nu)\leq0.1$. Thus the $R_2$ LQ model
becomes disfavored to explain the $R(D^{(*)})$ anomalies. The reason is that, the $R_2$ LQ model generates only $S_2$ and $T$ operators. At $m_b$ scale,
we have $C_{S_2}(m_b)\approx8.1C_T(m_b)$, which means $S_2$ operator contributes dominantly. Similar to the pure $S_2$ scenario, to explain the anomalies, a larger $\mathcal{B}(B_c\rightarrow\tau\nu)$ is induced thus the $R_2$ LQ model becomes disfavored.

Future measurements may also be helpful to test different LQ models, and we list the predictions in Table~\ref{tab:obser2} and Table \ref{tab:obser21}. For $S_1$ and $U_1$ LQ models, the predictions on $P_{\tau}(D)$ are a bit smaller than those in the SM, though with large uncertainties. For $P_{\tau}(D^*)$, $F_L^{D^*}$, and $A_{FB}(D,D^*)$, predictions in these two models are close to those in the SM.

Since the $R_2$ LQ model is already disfavored by the constraint $\mathcal B(B_c\rightarrow\tau\nu)<10\%$, we do not discuss its predictions here. However, it is important to point out that the model itself is not finally excluded. One reason is that the $\chi^2$ of $R_2$ is still close to the $95\%$ C.L. exclusion limit, and more importantly, in the previous estimation on $\mathcal{B}(B_c\rightarrow\tau\nu)$, the SM value of $\mathcal{B}(B_c\rightarrow J/\psi\mu\nu)\simeq(2\pm0.5)\%$ has been used, and the upper limit of $\mathcal{B}(B_c\rightarrow\tau\nu)$ is proportional to the value of $\mathcal{B}(B_c\rightarrow J/\psi\mu\nu)$~\cite{Akeroyd:2017mhr,Aaij:2019ths}. However $\mathcal{B}(B_c\rightarrow J/\psi\mu\nu)$ has not been measured and the SM prediction has large uncertainties~\cite{Bardhan:2019ljo}. If $\mathcal{B}(B_c\rightarrow J/\psi\mu\nu)$ is increased in future measurements/estimates, the upper limit on $\mathcal{B}(B_c\rightarrow\tau\nu)$ will become weaker. Thus future precision measurements on $\mathcal{B}(B_c\rightarrow J/\psi\mu\nu)$ or directly on $\mathcal{B}(B_c\rightarrow\tau\nu)$ will be very helpful for further testing the $R_2$ LQ model.
\begin{table*}[!htbp]\small%
\centering
\caption{Predictions for $R(D)$, $R(D^*)$, $P_\tau(D)$ and $P_\tau(D^*)$ by the leptoquark models. The first and second uncertainties are respectively due to the input parameters and the best-fit leptoquark couplings.}
\begin{tabular}{ccccc}
\hline
\hline
LQ type & $R(D)$& $R(D^*)$   &$P_\tau(D)$ & $P_\tau(D^*)$ \\
\hline
$S_1$   &$0.330(5)(30)$ & $0.301(10)(13)$ &$0.193(5)(147)$ &$-0.474(7)(20)$ \\
$U_1$   &$0.338(6)(30)$ & $0.298(10)(12)$ &$0.268(3)(88)$ &$-0.502(4)(16)$ \\
\hline
\hline
\end{tabular}
\label{tab:obser2}
\end{table*}
\begin{table*}[!htbp]\small%
\centering
\caption{Predictions for $F_L^{D^*}$, $\mathcal A_{FB}(D)$ and $A_{FB}(D^*)$ by the leptoquark models. The first and second uncertainties are respectively due to the input parameters and the best-fit leptoquark couplings.}
\begin{tabular}{cccc}
\hline
\hline
LQ type &$F_L^{D^*}$ &$\mathcal A_{FB}(D)$& $A_{FB}(D^*)$\\
\hline
$S_1$   &$0.480(4)(13)$ & 0.375(1)(12)& -0.061(5)(5)\\
$U_1$    &$0.463(4)(6)$ & 0.365(1)(6)& -0.063(6)(9)\\
\hline
\hline
\end{tabular}
\label{tab:obser21}
\end{table*}
\section{Summary and conclusions}
\label{sec:con}
In this work, we have performed combined analysis of the measurements of the LFU ratios $R(D^{(*)})$ and $R(J/\psi)$, the longitudinal polarizations $P_\tau(D^*)$ and $F_L^{D^*}$, and the branching ratio of the pure leptonic decay $B_c\to\tau\nu$. We have mainly studied two nontrivial aspects: the determination of the hadronic form factors and the new physics analysis.

For the $B\to D^{(*)}$ form factors, we have fitted the parameters in the HQET parametrization including the $\mathcal{O}(\alpha_s,\Lambda_{\mathrm{QCD}}/m_{b,c})$ corrections and part of $\mathcal{O}(\varepsilon_c^2)$ corrections. In the fit, we have taken into account the most recent theoretical results calculated using complementary QCD-based methods, lattice QCD and QCD light-cone sum rules, and we have also imposed the strong unitarity bounds that follow from the analytic properties of the QCD correlator, the quark-hadron duality and the crossing symmetry. We have obtained the optimal values for the parameters in the HQET parametrization with $\chi^2/d.o.f=22.30/23$. Using these HQET parameters, we have further converted strong unitarity bounds to constraints on individual helicity amplitudes, which have been used to check the consistency of our fit and can be used elsewhere in the BGL fit without imposing the HQET relations between form factors. Moreover, we have obtained the SM predictions for $R(D^{(*)})$ as well as other polarized and angular observables. Our results of $R(D^{(*)})$  are consistent with some recent predictions within the errors and deviate from the experimental averages at $\sim 4\sigma$.

We have also used the fitted form factors in the new physics study of the $b\to c\tau\nu$ anomalies, including the model-independent analysis and the study of LQ models. In the model-independent analysis, we have considered one-operator scenarios with complex Wilson coefficients and two-operator scenarios with real Wilson coefficients. We have found that the global fits for most scenarios give $\chi^2_{min}$ allowed within $95\%$ C.L., except the pure scalar scenarios $S_1$, $S_2$ and ($S_1$,$S_2$) (even with complex Wilson coefficients). These scalar scenarios give unacceptably large $\chi^2$ therefore the NP models that generate only scalar operators are ruled out, such as the charged Higgs models.

We have also studied three types of LQ models, namely $R_2$, $S_1$ and $U_1$ LQ models, which have been considered as possible explanations of the charged-current B anomalies in existing literatures. Our analysis has shown that the $S_1$ and the $U_1$ LQ models are still able to accommodate the current data, but the $R_2$ LQ model is already in tension with the limit $\mathcal{B}(B_c\to\tau\nu)<10\%$. Future measurements of $\mathcal{B}(B_c\to\tau\nu)$ or $\mathcal{B}(B_c\to J/\psi\mu\nu)$ will be very useful for testing the $R_2$ LQ model.

Using the best-fit Wilson coefficients, we have predicted various observables for the allowed NP scenarios and the LQ models. In addition to the observables considered in the global fits, we have also made predictions for the $\tau$ polarization $P_\tau(D)$ and the forward-backward asymmetries $A_{FB}(D)$ and $A_{FB}(D^*)$. Among these observables, the $\tau$ polarizations $P_\tau(D)$ have been found to be useful for separating the one-operator scenarios from the two-operator ones, while $P_\tau(D^*)$ and $A_{FB}(D^{(*)})$ could help further distinguish among the scenarios. In contrast, the $D^*$ longitudinal polarization has been found to be of little help in differentiating the NP scenarios because most scenarios give predictions close to the SM one. Moreover, all scenarios predict that $F_L^{D^*}$ is lower than 0.5, and therefore lower than the central value of the measurement. These observables are also helpful in testing $S_1$ and $U_1$ LQ models. In particular for $P_{\tau}(D)$, the predictions in both models are close to each other, and smaller than those in the SM, while for other observables, the predictions in both LQ models are close to the SM predictions. Future precision measurements of these observables at Belle II/LHCb~\cite{Kou:2018nap,Bediaga:2018lhg} will further help investigate NP effects in the $b\to c\tau\nu$ transition.

\section*{Acknowledgments}
The authors would like to thank Dante Bigi, Paolo Gambino, Martin Jung and Stefan Schacht for very useful discussions and communications on the $B\to D^{(*)}$ form factors. Z.R. Huang is grateful to Emi Kou for very helpful discussions on the topic. This work was supported in part by the MoST of Taiwan under the grant no.:107-2112-M-007-029-MY3 and the National Science Foundation of China under the grants 11847040, 11521505 and 11621131001.

\begin{appendix}
\section{Constraints on the Wilson coefficients for two-operator scenarios}
\label{app:A}
In this appendix, we present the plots that show the $2\sigma$ allowed regions by the various $b\to c\tau\nu$ measurements for each of the two-operator NP scenarios.
\begin{figure}[!h]
\begin{center}
\includegraphics[scale=0.22]{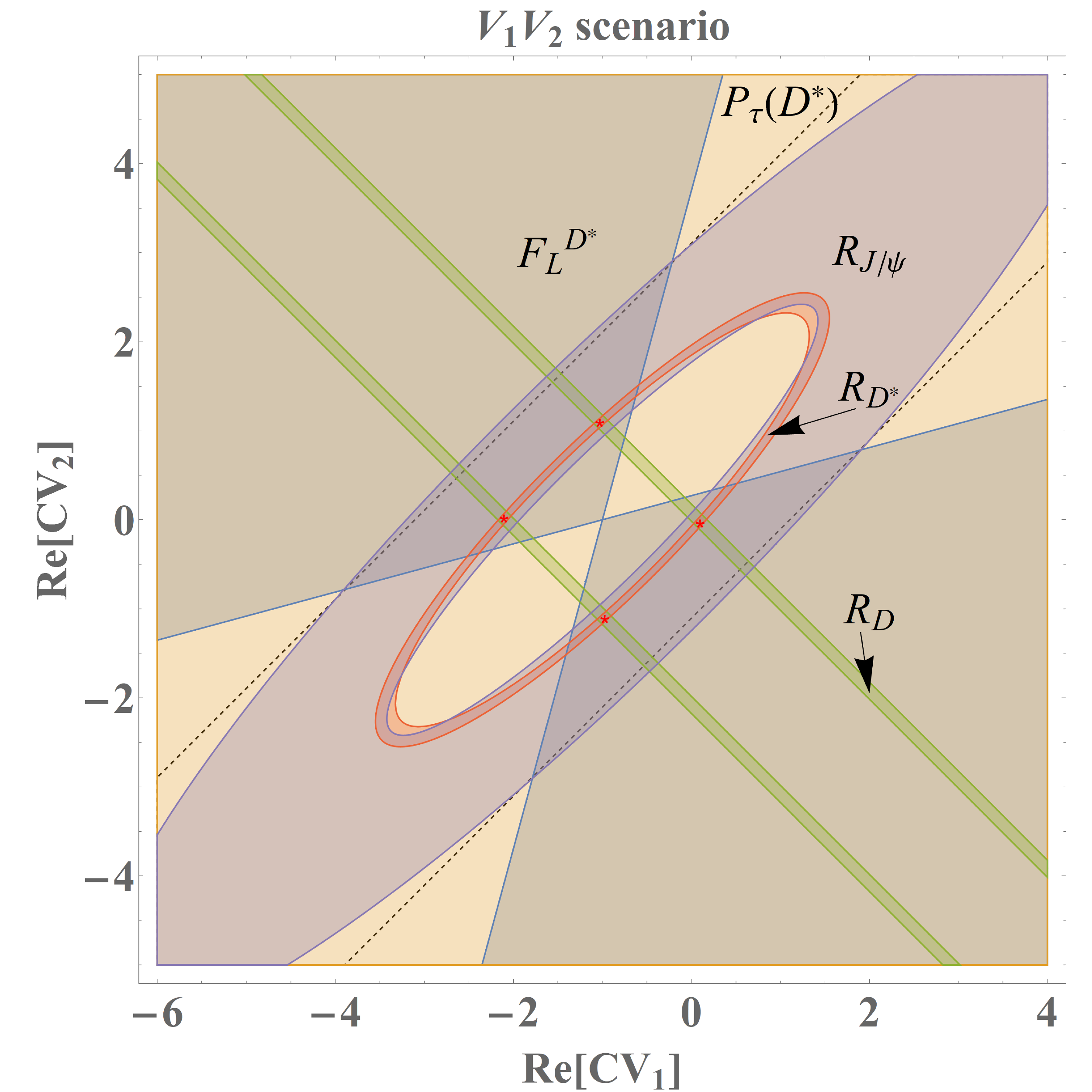}
\includegraphics[scale=0.22]{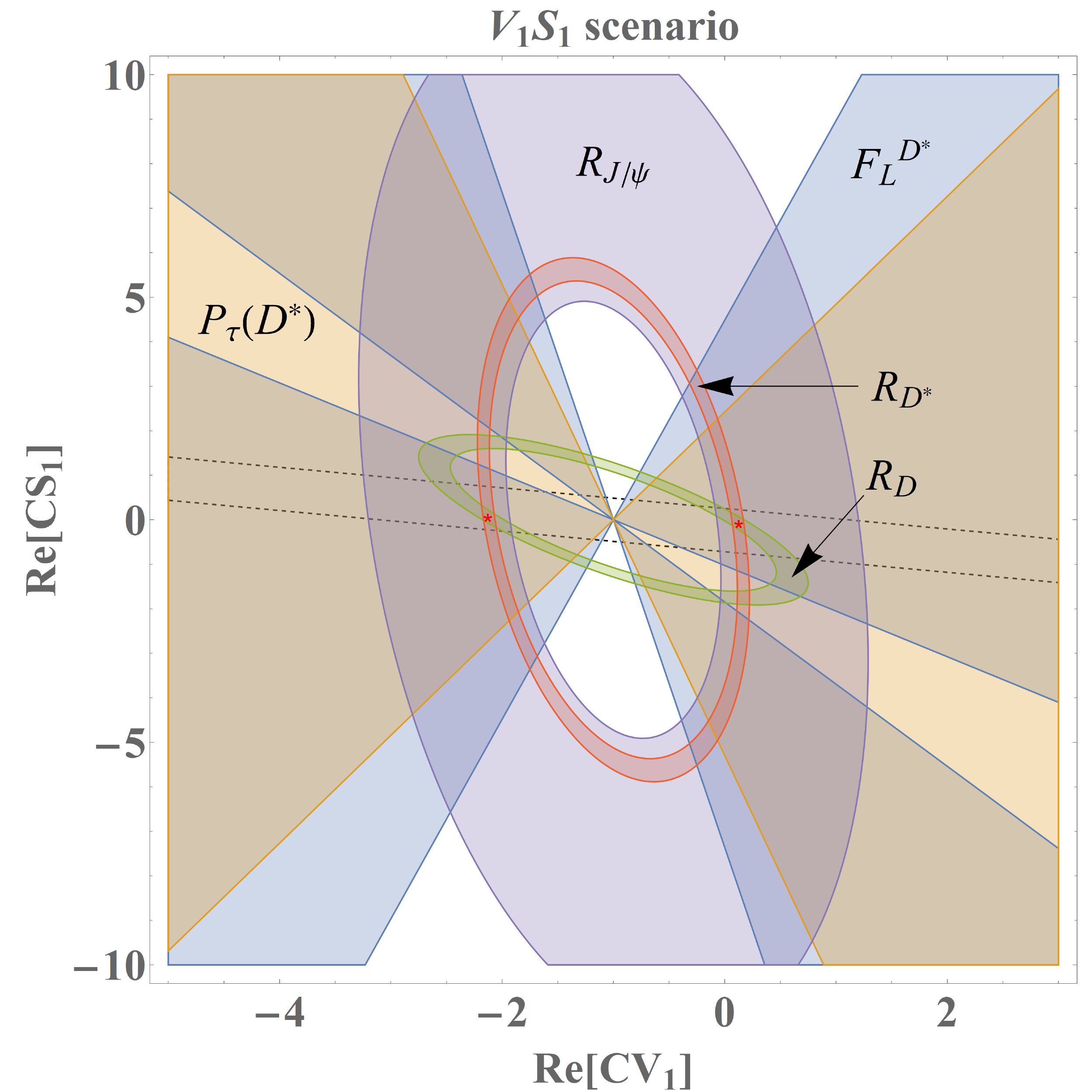}
\includegraphics[scale=0.22]{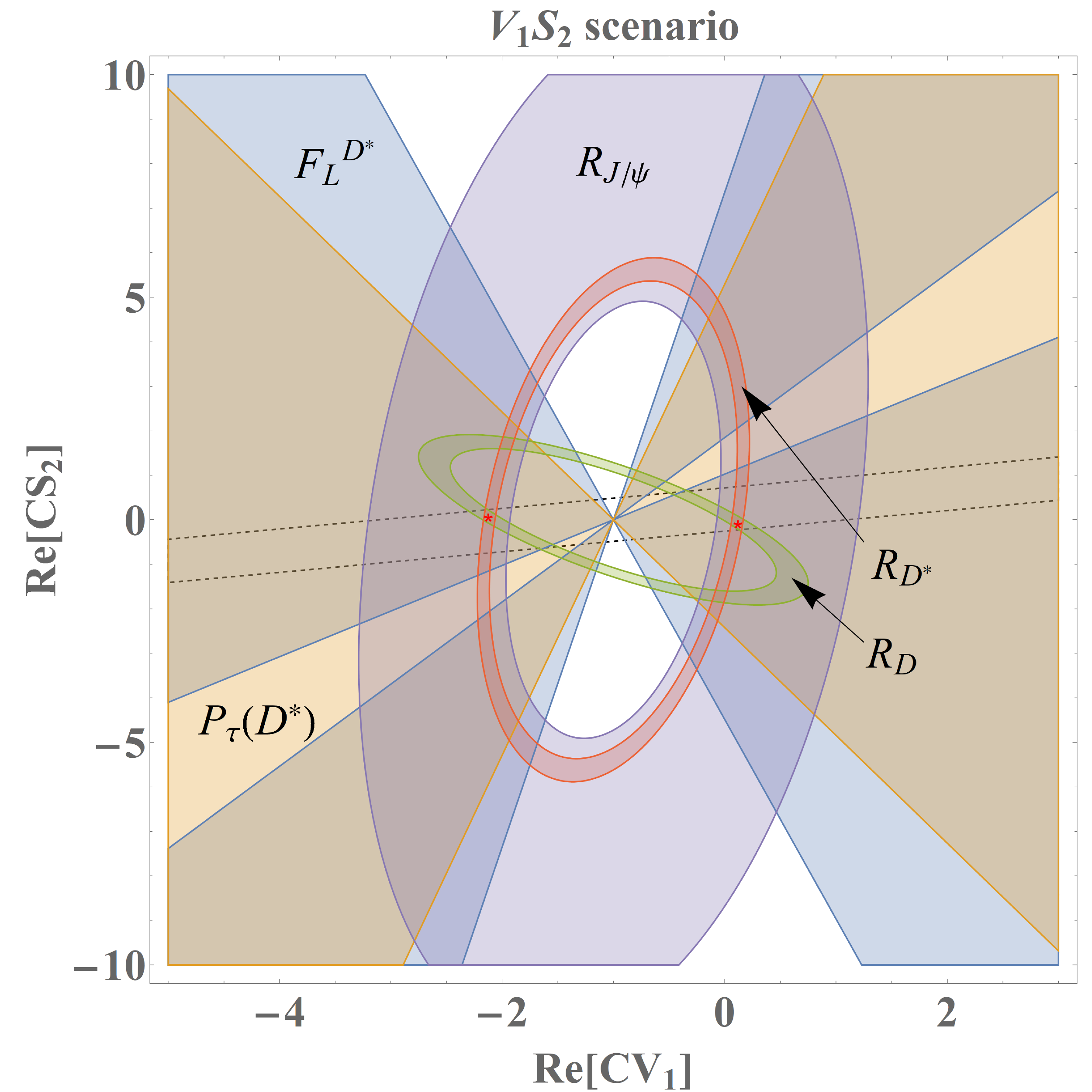}
\includegraphics[scale=0.22]{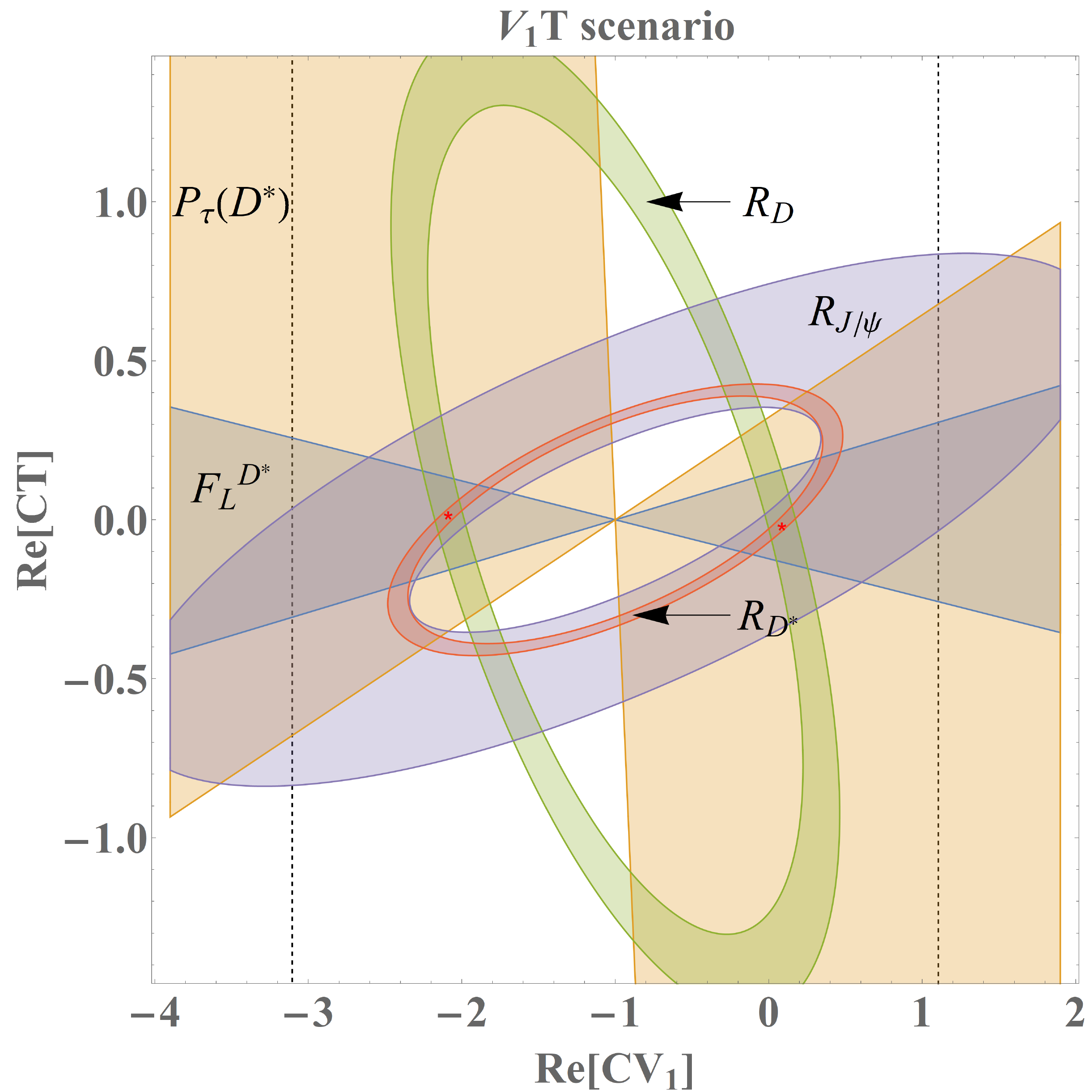}
\includegraphics[scale=0.22]{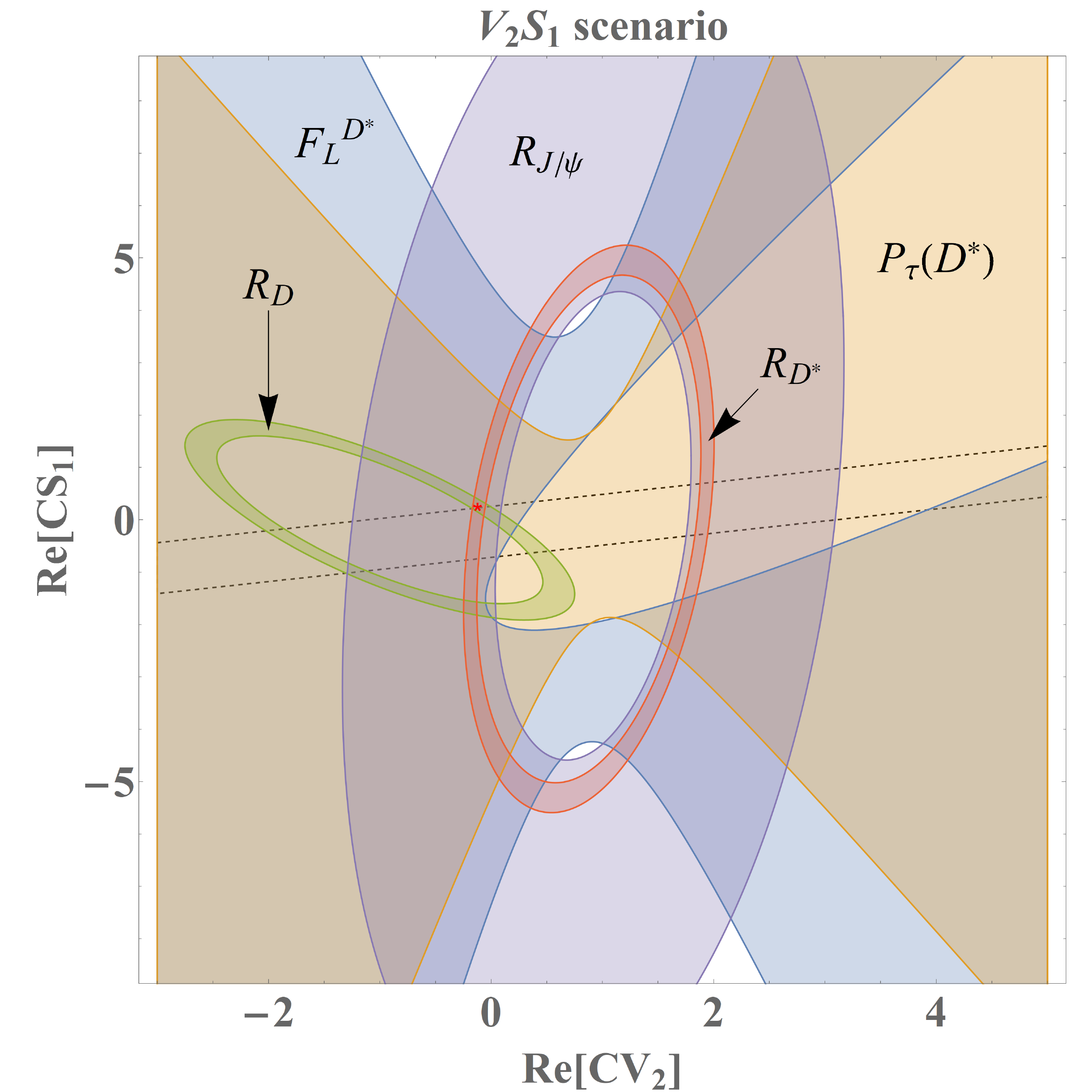}
\includegraphics[scale=0.22]{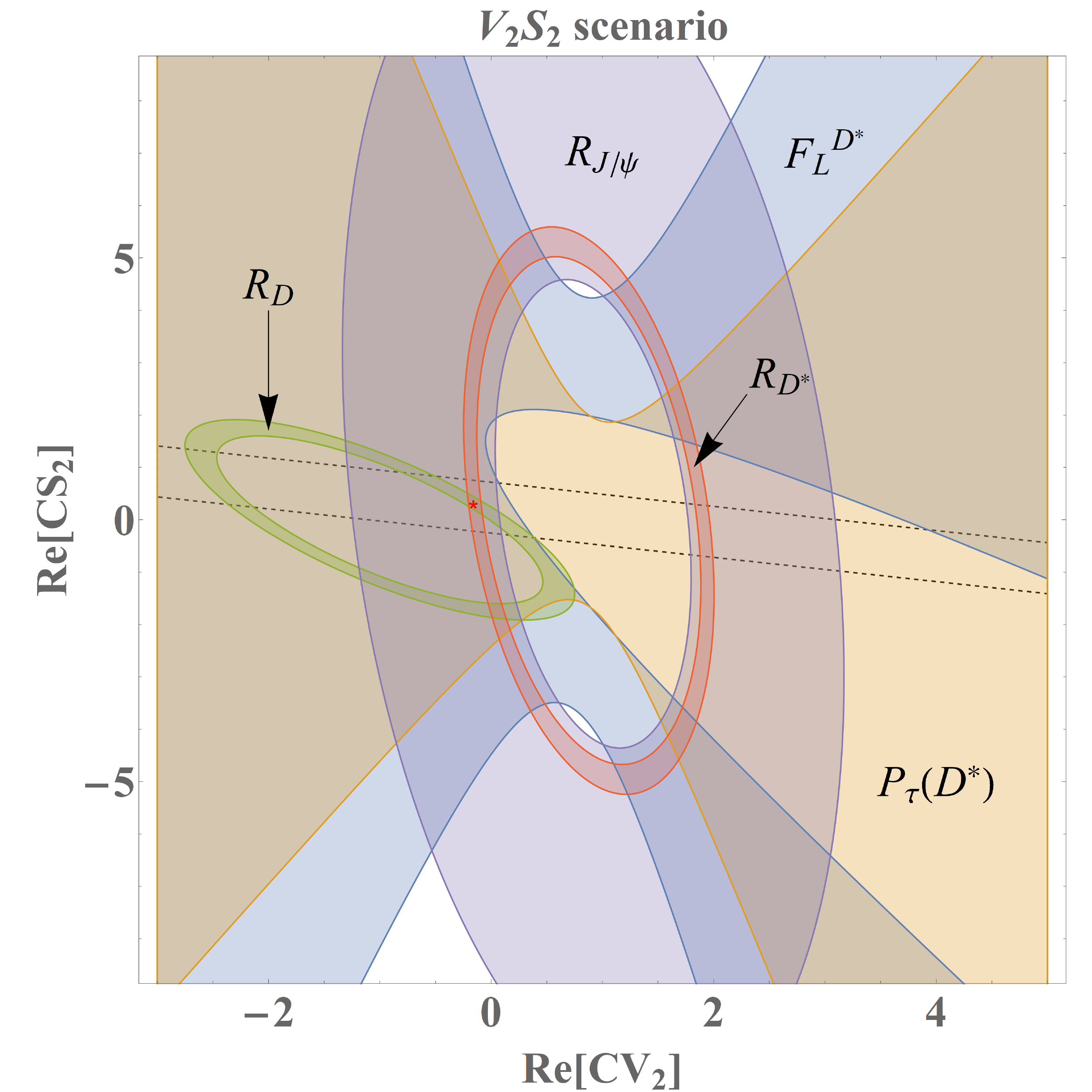}
\caption{
Constraints on the Wilson coefficients in two-operator NP scenarios by the measurements of $R(D^{(*)})$, $R(J/\psi)$, $P_{\tau}(D^*)$ and $F_L^{D^*}$ at $95\%$ C.L. and the limit on $\mathcal B(B_c \to \tau\nu)$ (black dashed curves). The red stars and red dashed curves denote the Wilson coefficients fitted without taking into account $\mathcal B(B_c \to \tau\nu)<10\%$.}
\label{fig:constraint2sce1}
\end{center}
\end{figure}
\begin{figure}[!h]
\begin{center}
\includegraphics[scale=0.22]{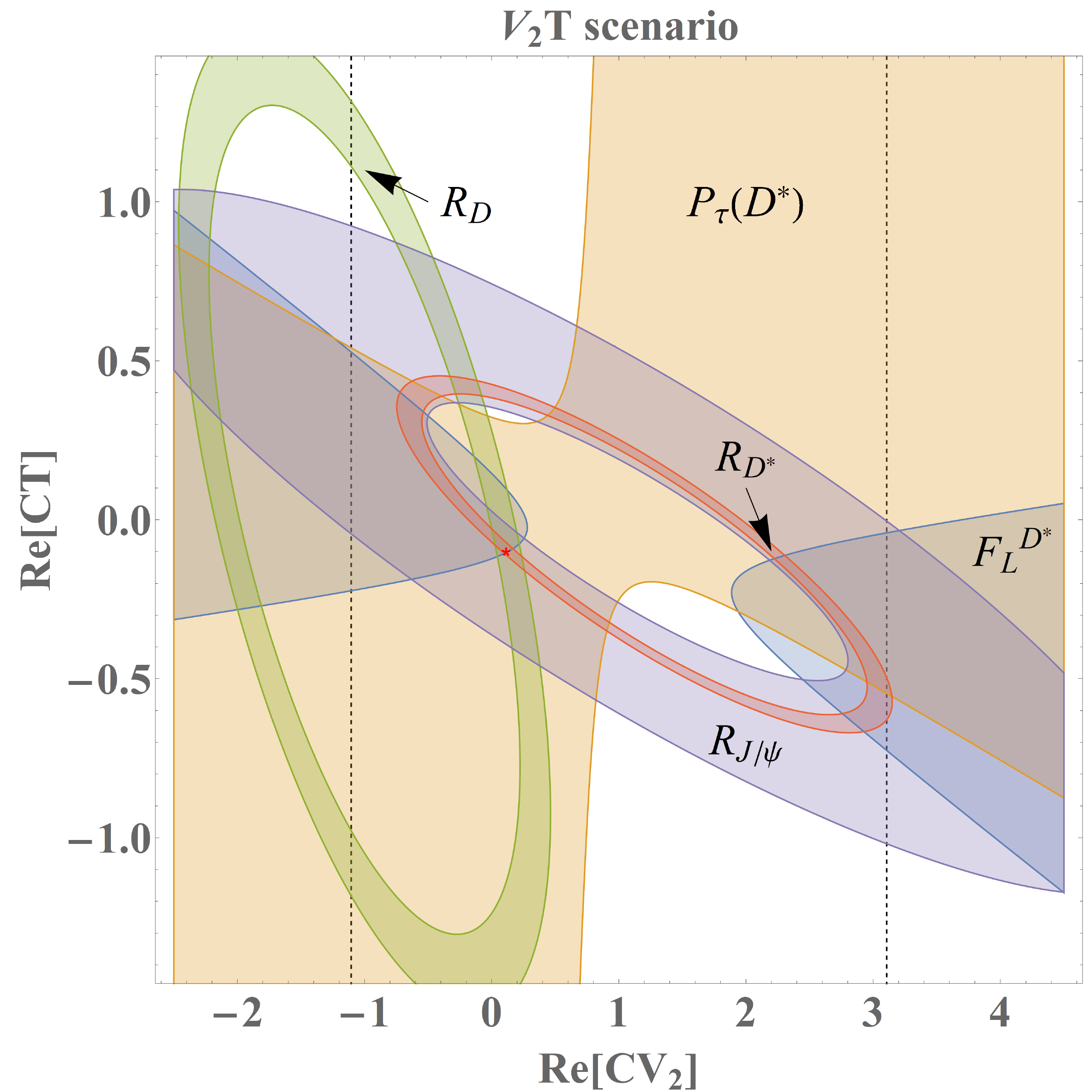}
\includegraphics[scale=0.22]{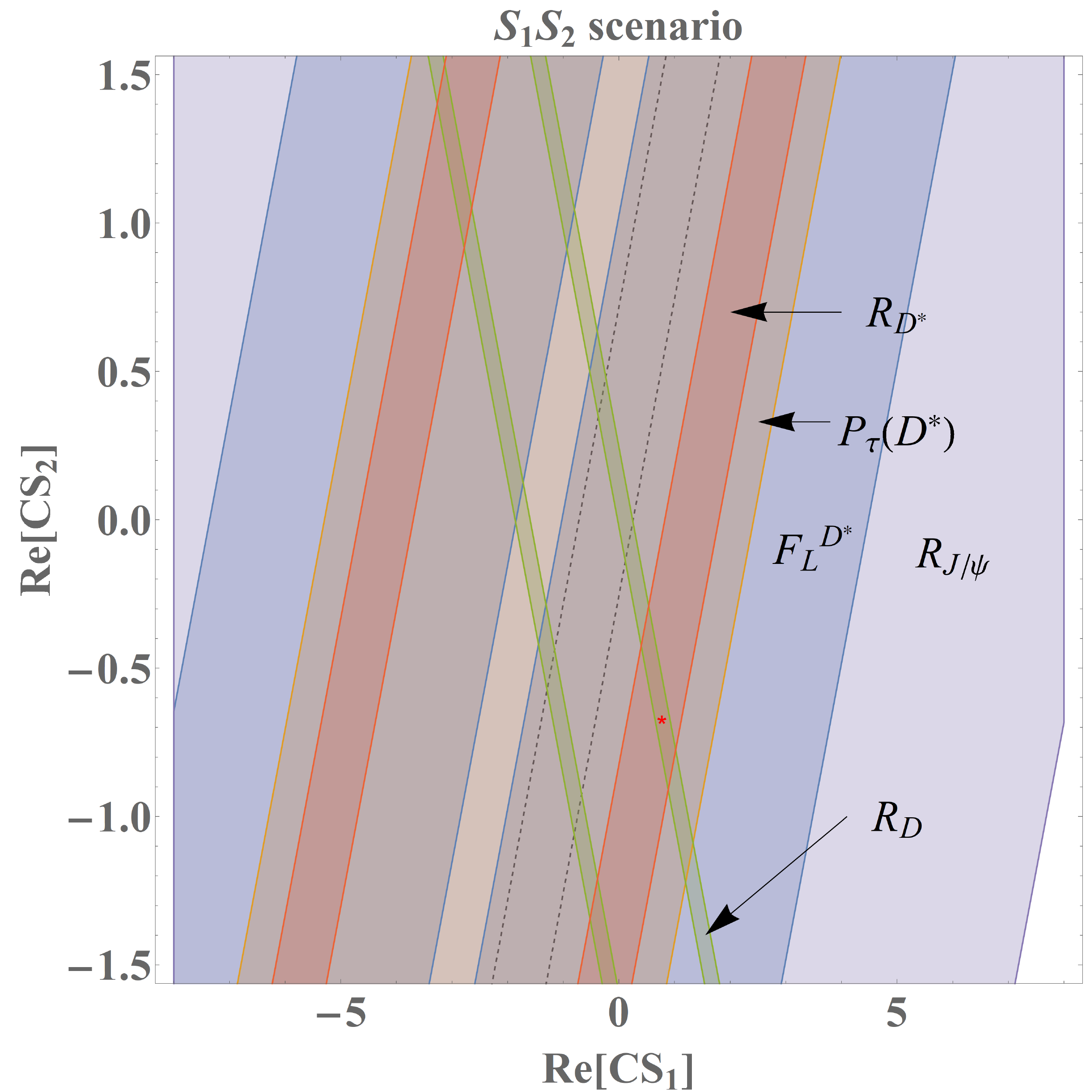}
\includegraphics[scale=0.22]{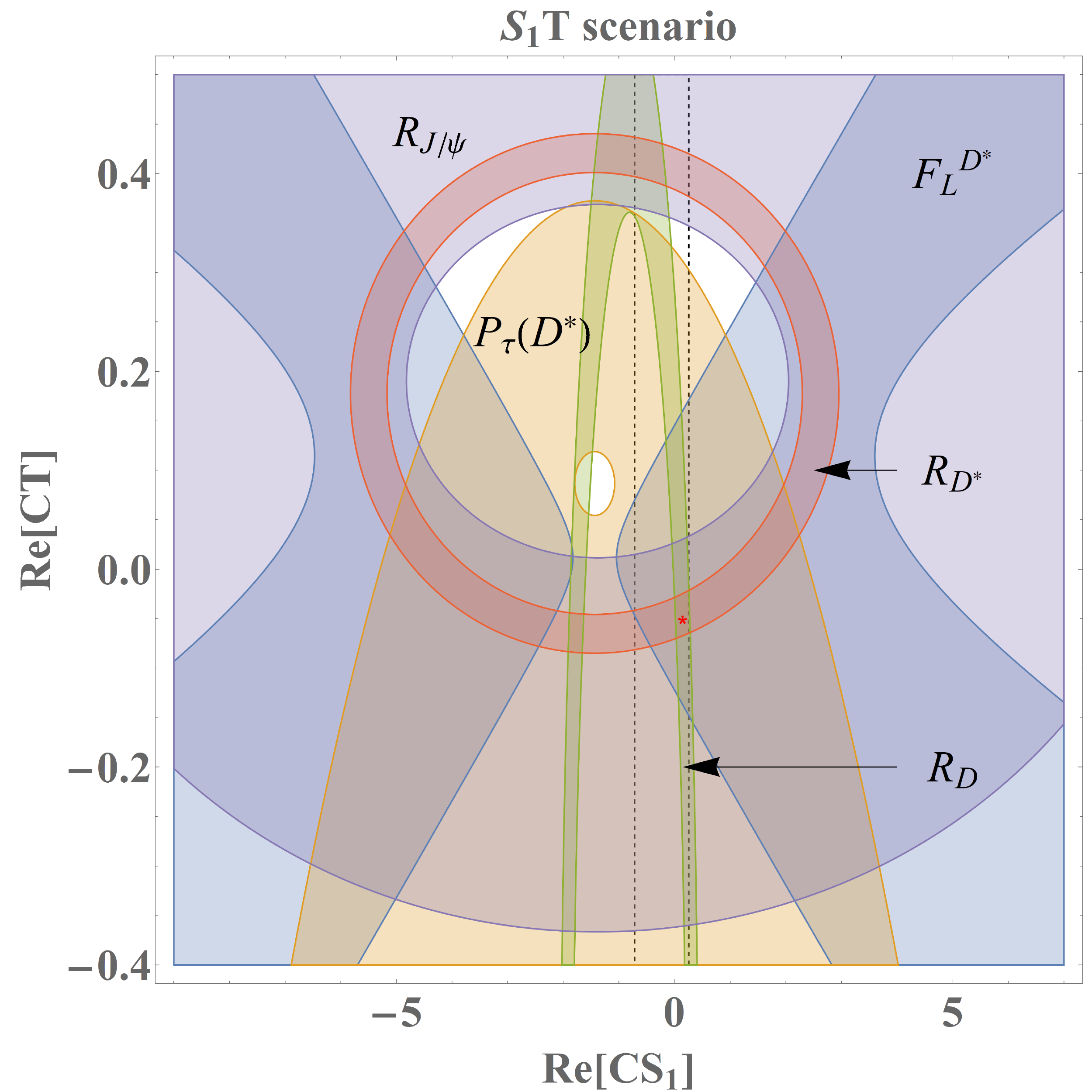}
\includegraphics[scale=0.22]{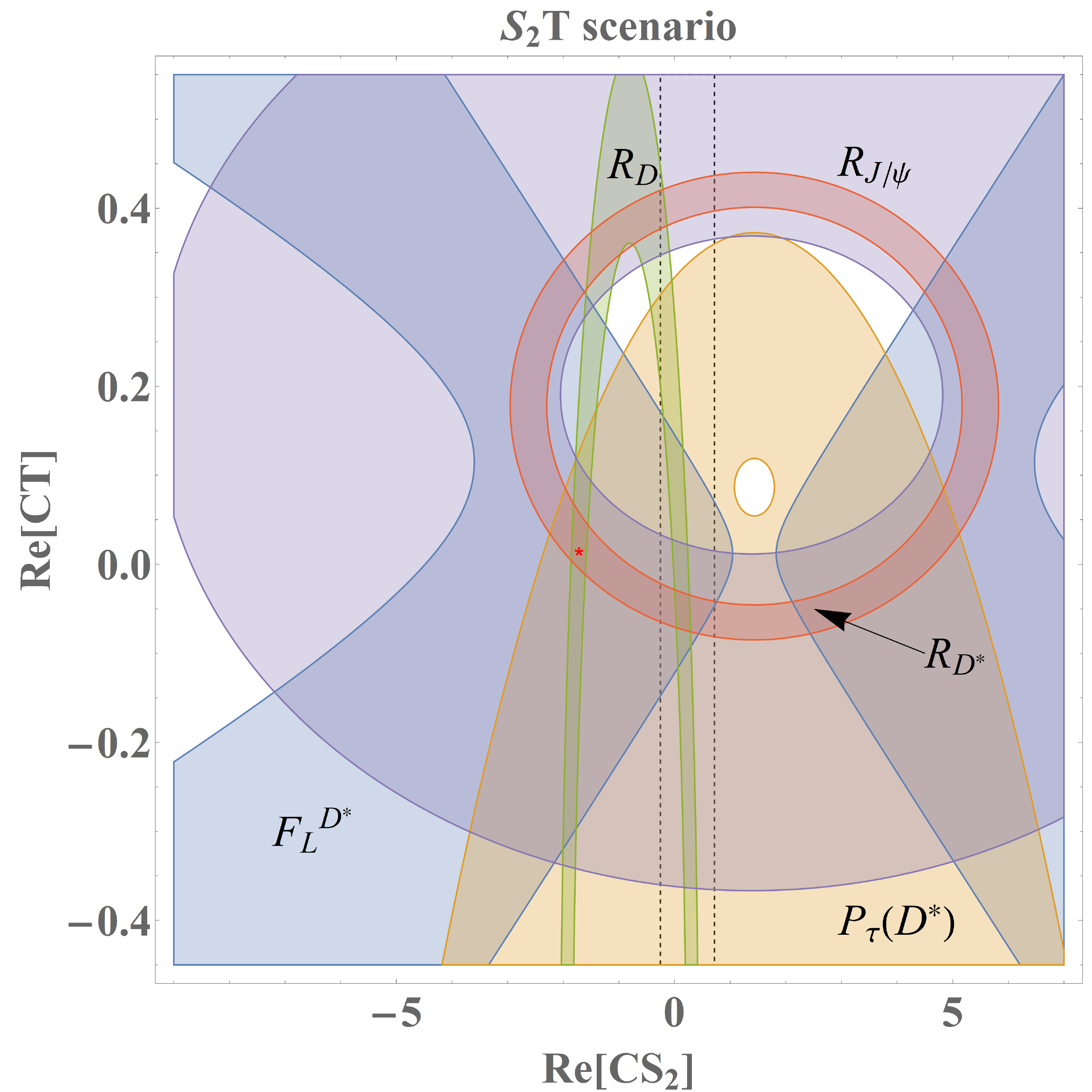}
\caption{
Constraints on the Wilson coefficients in two-operator NP scenarios by the measurements of $R(D^{(*)})$, $R(J/\psi)$, $P_{\tau}(D^*)$ and $F_L^{D^*}$ at $95\%$ C.L. and the limit on $\mathcal B(B_c \to \tau\nu)$ (black dashed curves). The red stars and the red dashed curves denote the Wilson coefficients fitted without taking into account $\mathcal B(B_c \to \tau\nu)<10\%$.}
\label{fig:constraint2sce2}
\end{center}
\end{figure}
\end{appendix}
\clearpage
\bibliography{myreference}
\end{document}